\documentclass[10pt,aps,prd,twocolumn,superscriptaddress,amsmath, natbib, nofootinbib,floats,floatfix]{revtex4-1}
\usepackage{bm}
\usepackage{amsmath}
\usepackage{amsfonts}
\usepackage{aas_macros}
\usepackage{graphicx}
\usepackage[hidelinks]{hyperref}
\usepackage{booktabs}
\usepackage{color}
\usepackage[dvipsnames]{xcolor}
\usepackage[capitalise]{cleveref}
\usepackage{subfigure}

\creflabelformat{equation}{#2\textup{#1}#3}

\begin{document}

\title{Constraints on Galileons from the positions of supermassive black holes}

\author{D. J. Bartlett}
\email{deaglan.bartlett@physics.ox.ac.uk}
\affiliation{Astrophysics, University of Oxford, Denys Wilkinson Building, Keble Road, Oxford, OX1 3RH, United Kingdom}
\author{H. Desmond}
\email{harry.desmond@physics.ox.ac.uk}
\affiliation{Astrophysics, University of Oxford, Denys Wilkinson Building, Keble Road, Oxford, OX1 3RH, United Kingdom}
\author{P. G. Ferreira}
\affiliation{Astrophysics, University of Oxford, Denys Wilkinson Building, Keble Road, Oxford, OX1 3RH, United Kingdom}

\begin{abstract}
    Galileons are scalar field theories which obey the Galileon symmetry $\varphi \to \varphi + b + c_\mu x^\mu$ and are capable of self-acceleration if they have an inverted sign for the kinetic term. These theories violate the Strong Equivalence Principle, such that black holes (BHs) do not couple to the Galileon field, whereas non-relativistic objects experience a fifth force with strength $\Delta G / G_{\rm N}$ relative to gravity. For galaxies falling down a gradient in the Galileon field, this results in an offset between the centre of the galaxy and its host supermassive BH. We reconstruct the local gravitational and Galileon fields through a suite of constrained N-body simulations (which we dub \texttt{CSiBORG}) and develop a Monte Carlo-based forward model for these offsets on a galaxy-by-galaxy basis. Using the measured offset between the optical centre and active galactic nucleus of 1916 galaxies from the literature, propagating uncertainties in the input quantities and marginalising over an empirical noise model describing astrophysical and observational noise, we constrain the Galileon coupling to be $\Delta G / G_{\rm N} < 0.16$ at $1\sigma$  confidence for Galileons with crossover scale $r_{\rm C} \gtrsim H_0^{-1}$.
\end{abstract}

\maketitle

\section{Introduction}

The end of the 20\textsuperscript{th} century saw the startling discovery that the Universe’s expansion is accelerating \cite{Riess_1998,Perlmutter_1999}. Since General Relativity (GR) has successfully reproduced a century's worth of observations across a wide range of environments \cite{Burrage_2018}, one potential explanation is that this acceleration is due to a negative pressure component of stress-energy: dark energy \cite{Frieman_2008}. Our current cosmological model ($\Lambda$CDM) assumes this takes the form of a cosmological constant,although this has issues with radiative instability and UV-sensitivity (the cosmological constant problem \cite{Padilla_2015}). This, coupled with tensions between competing probes of the Universe \cite{Cosmology_Intertwined_1,Cosmology_Intertwined_2,Cosmology_Intertwined_3}, leads us to consider an alternative explanation, namely that GR may not be the correct description of gravity on large scales.

Modified gravity theories generically violate the Strong Equivalence Principle (SEP), which refers to the equivalence of free fall independent of an object's gravitational binding energy or composition
\cite{Joyce_2016}. Despite little success so far in explaining cosmic acceleration \cite{Novel_Probes_2019}, studies of modified gravity are essential as they provide consistent, plausible alternatives to GR and could relieve some of the current tensions (e.g. \cite{Sola_2019,Frusciante_2020, Desmond_H0, Desmond_H0_2}).

On astrophysical scales, modified gravity theories lead to a fifth fundamental force \cite{Jain_2011}. This force is not detected in laboratory or Solar System experiments, so must be `screened' to remain hidden in these environments.
The mechanism through which this is achieved falls into one of three categories (see \cite{Jain_2010,Joyce_2015,Khoury_2010,Novel_Probes_2019} for reviews of screened modified gravity theories). The `thin-shell' mechanism (e.g. chameleon \cite{Khoury_2004a,Khoury_2004b}, symmetron \cite{Hinterbichler_2010} and dilaton \cite{Brax_2010}) is determined by the gravitational potential, whereas `kinetic' screening (e.g. K-mouflage \cite{Babichev_2009} and Vainshtein \cite{Vainshtein_1972}) depends on derivatives of the potential. The third mechanism depends on interactions between baryons and dark matter such that the gravitational constant depends on the local dark matter density \cite{Sakstein_2019}.

Thin-shell screened theories are relatively well-constrained: the chameleon mechanism has been probed with a wide variety of astrophysical and laboratory signals \cite{Burrage_2016}, and recently \cite{f(R)_ruled_out} astrophysically relevant $f(R)$ theories have been ruled out by galaxy morphology. Vainshtein-screened theories however have many fewer tests, leading us to focus on them here. The observational test we consider was introduced by \citet{Hui_2012}, and is based on the same physical principle as the test presented in \citet{Sakstein_2017} and \citet{Asvathaman_2017}: we investigate whether the offset between a galaxy's centre and its central supermassive black hole (BH) is preferentially aligned along the local gravitational field, as would be expected by the SEP violation in this class of modified gravity theories.

The success of Monte Carlo-based forward models to constrain fundamental physics in the astrophysical regime \cite{Desmond_2018,Desmond_2018_warp,Pardo_2019, Desmond_letter,f(R)_ruled_out} motivates us to adapt this approach for the BH test of Galileons. Specifically, we forward model the magnitude and direction of the BH--galaxy offset for each galaxy in the samples collated in \cite{Bartlett_2020}. The measured BH--galaxy offset for these galaxies is determined by cross-matching the optical centre to observations at a different wavelength, which provides the location of the active galactic nucleus.
We introduce a suite of constrained N-body simulations of the local universe (\texttt{CSiBORG}), which we use to map out the large-scale Galileon field. Combining this with models for galaxy and halo structure allows us to make predictions for the BH--galaxy offsets. Marginalising over uncertainties in the Galileon field and galaxy properties, as well as parameters describing the noise due to non-fifth-force contributions to the signal, we compare our predictions to the observed offsets via a Markov Chain Monte Carlo (MCMC) algorithm. We find that the strength of the fifth force relative to gravity, $\Delta G / G_{\rm N}$, is constrained to be $<0.16$ at $1 \sigma$ confidence; this bound is applicable to Galileons with crossover scales $r_{\rm C} \gtrsim H_0^{-1}$.

In \Cref{sec:Theory} we discuss this phenomenon in the context of the cubic Galileon model, although the effect occurs more generally in Galileon theories due to their BH no hair theorem \cite{Hui_2013}. We present the observational data used in this work in \Cref{sec:Observational data}. \Cref{sec:Methods} details our inference methods and the results are presented in \Cref{sec:Results}. We discuss systematic uncertainties and compare our constraints to previous work in \Cref{sec:Discussion} and conclude in \Cref{sec:Conclusions}.

\section{Galaxy--black hole offsets in Galileon gravity}
\label{sec:Theory}

We consider a theory containing a single scalar field, $\varphi$, which respects the Galileon symmetry $\varphi \to \varphi + b + c_\mu x^\mu$ \cite{Nicolis_2009} and is Vainshtein screened \cite{Vainshtein_1972} on small scales. A common example is the cubic Galileon, which has the action
\begin{equation}
    S = \int {\rm d}^4 x \sqrt{-g} \left[ \frac{R}{16 \pi G} - \frac{1}{2} c_2 \mathcal{L}_2 - \frac{1}{2} c_3 \mathcal{L}_3 - \mathcal{L}_{\rm m} \right],
\end{equation}
where $R$ is the Ricci scalar, $g$ is the determinant of the metric $g_{\mu\nu}$, $c_3$ and $c_4$ are constants, $\mathcal{L}_{\rm m}$ is the matter Lagrangian, and
\begin{equation}
    \mathcal{L}_2 = \nabla_\mu \varphi \nabla^\mu \varphi, \quad \mathcal{L}_3 = \frac{2}{\mathcal{M}^3} \square \varphi \nabla_\mu \varphi \nabla^\mu \varphi,
\end{equation}
where $\mathcal{M}^3 = M_{\rm Pl}H_0^2$. There are two branches to the cubic Galileon, depending on the sign of the kinetic term. If, using the mostly minus signature, $c_2 > 0$ (normal branch) we have a scalar field with a canonical kinetic term; these cannot self-accelerate and are simply models of fifth-forces. On the other hand, if $c_2 < 0$ (self-accelerating branch) then the field can self-accelerate and does not necessarily require a cosmological constant \cite{Babichev_2013}. 

If we work in the quasi-static approximation and neglect terms suppressed by the Newtonian potentials and their spatial derivatives, we obtain \cite{Barreria_2013} the equation of motion for perturbations, $\varphi$, about a background, $\bar{\varphi}$,
\begin{equation}
    \label{eq:cubic Galileon Barreira}
    \begin{split}
        \nabla^2 \varphi + \frac{1}{3 \beta_1 a^2 \mathcal{M}^3} &\left[ \left( \nabla^2 \varphi \right)^2 - \nabla_i \nabla_j \varphi \nabla^i \nabla^j \varphi \right] \\
        &= \frac{M_{\rm Pl}}{3 \beta_2} 8 \pi G_{\rm N} a^2 \bar{\rho} \Delta,
    \end{split}
\end{equation}
where $i \in \{1,2,3\}$ and
\begin{align}
    \beta_1 &= \frac{1}{6c_3} \left[ - c_2 - \frac{4 c_3}{\mathcal{M}^3} \left( \ddot{\bar{\varphi}} + 2 H \dot{\bar{\varphi}}\right)  + \frac{2 \kappa c_3^2}{\mathcal{M}^6} \dot{\bar{\varphi}}^4 \right], \label{eq:beta1 definition}\\
    \beta_2 &= 2 \frac{\mathcal{M}^3M_{\rm Pl}}{\dot{\bar{\varphi}}^2} \beta_1, \label{eq:beta2 definition}
\end{align}
with $\kappa \equiv 8 \pi G$. Our test will depend only on \autoref{eq:cubic Galileon Barreira} which holds for both signs of $c_2$; our constraints therefore apply to both branches. We can rewrite \autoref{eq:cubic Galileon Barreira} in a more familiar form,
\begin{equation}
    \label{eq:cubic Galileon}
    \nabla^2 \varphi + \frac{r_{\rm C}^2}{3} \left[ \left( \nabla^2 \varphi \right)^2 - \nabla_i \nabla_j \varphi \nabla^i \nabla^j \varphi \right] = 8 \pi \alpha G_{\rm N} \bar{\rho} \Delta,
\end{equation}
where $\alpha$ describes the strength of the coupling of the Galileon to matter, and $r_{\rm C}$, called the `crossover scale', parametrises the new kinetic terms.
We note that, using \Cref{eq:beta1 definition,eq:beta2 definition}, these parameters are functions of time. In this work we use low-redshift observations so will ignore this temporal evolution and consider only their present day values. Since the coupling in self-accelerating models tends to dramatically increase as we approach the present day \cite{Barreria_2013}, the constraints we find imply a bound on $\alpha$ over the history of the Universe for this branch.

To remain agnostic to the details of the Galileon model, we assume that $\alpha$ and $r_{\rm C}$ are independent. This is not true for all models: in the Dvali-Gabadadze-Porrati (DGP) model \cite{DGP_2000}, for example, $\alpha$ is related to $r_{\rm C}$ and the Hubble parameter, $H(t)$, as \cite{Koyama_2007}
\begin{equation}
    \label{eq:alpha DGP}
    \alpha_{\rm DGP} \left( t \right) = \frac{1}{3} \left[ 1 \pm 2 H r_{\rm C} \left( 1 + \frac{\dot{H}}{3 H^2} \right) \right]^{-1}.
\end{equation}
An overdot denotes a derivative with respect to cosmic time $t$, the $+$ sign refers to the normal branch and the $-$ sign to the self-accelerating branch. To test a specific model, one should compare the model's trajectory in the $\alpha-r_{\rm C}$ plane to the constraints obtained in this work.

Far outside the Vainshtein radius, $r_{\rm V}$, the new kinetic terms are negligible and we recover Poisson's equation. For a source of mass $M$, this transition occurs at
\begin{equation}
    \label{eq:Vainshtein radius definition}
    r_{\rm V} = \left( \frac{4}{3} \alpha G_{\rm N} Mr_{\rm C}^2 \right)^{\frac{1}{3}}.
\end{equation}
Within the Vainshtein radius the fifth force has magnitude
\begin{equation}
    \label{eq:fifth force for Galileon}
    a_{5} = -\alpha \nabla \varphi = \frac{\Delta G}{G_{\rm N}} \frac{G_{\rm N} Q M}{r^2} \left( \frac{r}{r_{\rm V}} \right)^q
\end{equation}
where $q = 3/2$ for the cubic Galileon, $Q$ is the scalar charge of the object given in terms of its stress-energy tensor as
\begin{equation}
    Q = \int T^0{}_0 {\rm d}^3 x,
\end{equation}
and
\begin{equation}
    \frac{\Delta G}{G_{\rm N}} \equiv 2\alpha^2.
\end{equation}
The suppression of $a_5$ for $r \ll r_V$ is what constitutes Vainshtein screening.

For a non-relativistic object, $Q$ is equivalent to the object's mass, but for compact objects $Q < m$ because $T$ does not include gravitational binding energy. The limiting case is a black hole, for which $Q = 0$.

Due to the Galileon symmetry, by adding a term with a linear gradient, one can always generate a new solution $\varphi \to \varphi + \varphi_{\rm ext}$, where $\nabla\varphi_{\rm ext}$ is a constant. Therefore, although stars in galaxies tend to reside within their host's Vainshtein radius, this does not mean they cannot feel a fifth force. Rather, they interact with the field sourced by large scale structure \cite{Hui_2012}, which has a wavelength long compared to the Vainshtein radius and hence has approximately constant gradient on the scale of the galaxy. Cosmological simulations have confirmed this prediction \cite{Falck_2014} and indicate that $\varphi$ obeys linear dynamics on scales $\gtrsim 10 {\rm \, Mpc}$ for $r_{\rm C} \simeq 6 {\rm \, Gpc}$ \cite{Cardoso_2008,Schmidt_2009,Khoury_2009,Chan_2009}.

In conjunction with the no-hair theorem described above, this property of the Galileon symmetry can lead to an offset between the centre of a galaxy and its central BH. This occurs since, if a galaxy is falling down a scalar field potential, the non-relativistic matter feels the attractive fifth force, whereas the BH does not ($Q=0$). Therefore the BH lags behind the galaxy. The offset is stabilised by the gravitational force between the BH and the galaxy and its dark matter halo, which can lead to a constant displacement in equilibrium.

\section{Observational data}
\label{sec:Observational data}

In this work we use the four largest datasets collated and summarised in \cite{Bartlett_2020} which contain measurements of the offsets between an active galactic nucleus (AGN) and its host galaxy's centre: OF13 \cite{Orosz_2013}, O16 \cite{Orosz_2016}, SB18 \cite{Skipper_2018} and B19 \cite{Barrows_2019}. Each of these cross-match the optical centres of galaxies from the Sloan Digital Sky Survey (SDSS) \citep{Alam_2015} to observations at a different wavelength, where the latter provides the position of the AGN. OF13, O16 and SB18 search for radio counterparts, using the International Celestial Reference Frame (ICRF2) \cite{Fey_2015}, mJIVE-20 \cite{Deller_2014} and the Cosmic Lens All-Sky Survey (CLASS) \cite{Myers_2003,Browne_2003} respectively. In B19 AGN positions are obtained from \textit{Chandra} X-ray data \cite{Evans_2010}. The distributions of offsets from these samples are dominated by a Gaussian component describing the spatial resolution of the measurements, with width $\sigma_{\rm obs} \sim 50 {\rm \, mas}$ for the radio samples and $\sigma_{\rm obs} \sim 150 {\rm \, mas}$ for B19. Approximately 10-30 per cent of the probability density can be attributed to a non-Gaussian component (e.g. a Laplace distribution) which is dominant in the tails of the distribution \cite{Bartlett_2020}. The degree to which an AGN is intrinsically offset is given by the ratio of the non-Gaussian to Gaussian terms. 

\begin{figure}
	\centering
	\includegraphics[width=\columnwidth]{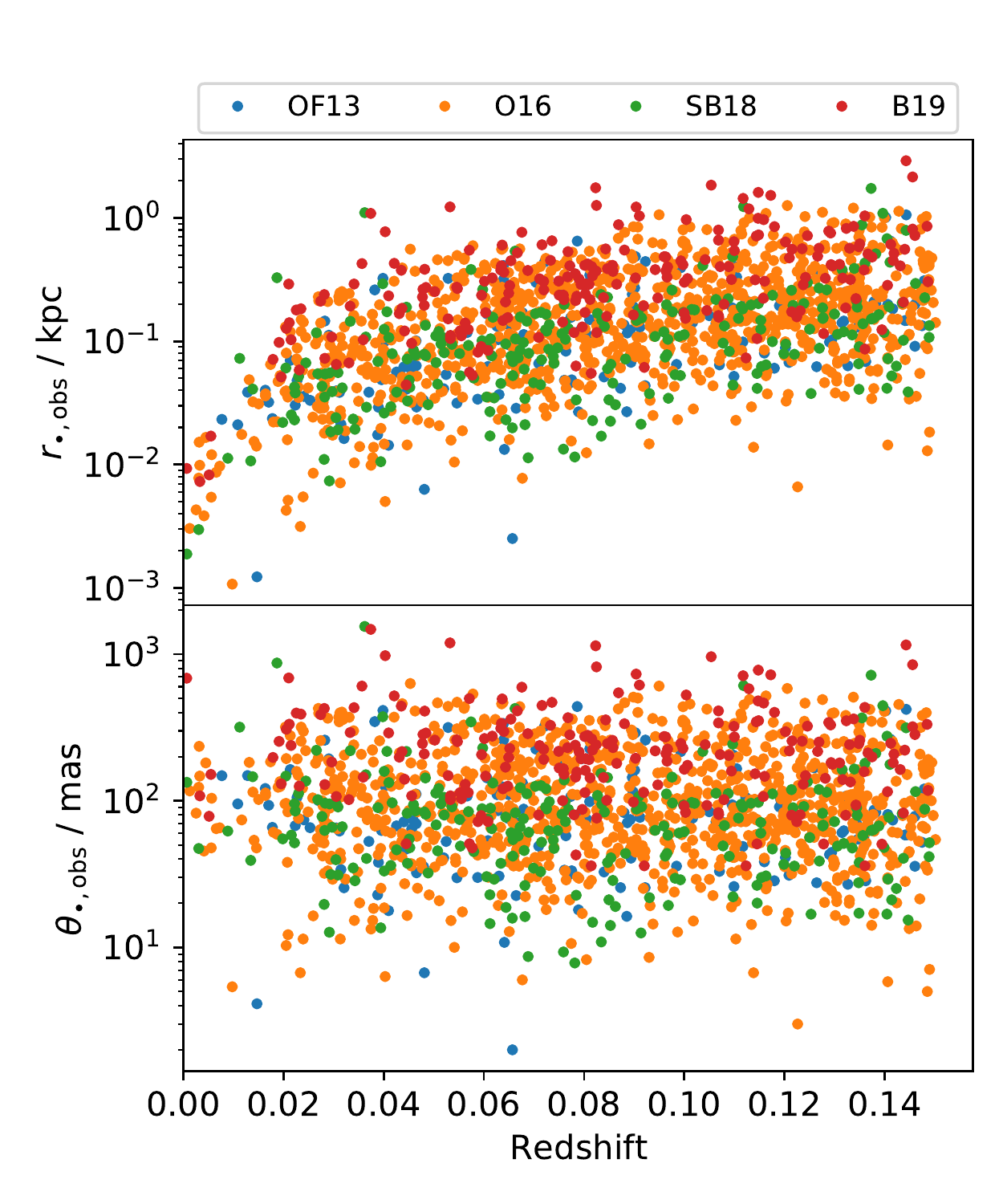}
	\caption{\label{fig:offset_data}Observed BH--galaxy offset as a function of redshift for the galaxies used in this work. The upper panel gives the physical offsets and the low panel gives the angular offsets. For redshift we use \texttt{zdist} from the NSA.}
\end{figure}

We plot the physical and angular offsets as a function of redshift for the galaxies used in this work in \autoref{fig:offset_data}. The galaxies are typically at redshift $z \sim 0.1$, such that a $3 \sigma_{\rm obs}$ offset for a galaxy from the radio samples corresponds to a physical offset of $\sim 340 {\rm \, pc}$.

For information on the halo structures (which determine the restoring force), we cross-correlate these data with the Nasa Sloan Atlas (NSA)\footnote{\url{ www.nsatlas.org}} to find the closest source within $0.5^\prime$. The NSA contains measured and derived quantities for nearby galaxies using state of the art sky subtraction and photometric determinations \cite{Blanton_2011} in the optical and near-infrared, largely from the Sloan Digital Sky Survey.
Approximately 10 per cent of galaxies are discarded due to not having an NSA counterpart, and we retain 144, 1328, 230 and 214 galaxies for the OF13, O16, SB18 and B19 samples respectively.

For distances, we use \texttt{zdist}, which is determined using the peculiar velocity model of \citet{Willick_1997}. Since, as described in \Cref{sec:Calculating the offset}, we use a S\'{e}rsic profile to determine the central baryonic surface density, we use quantities relevant to such a profile: stellar mass $M_{\star} = \texttt{sersic\_mass}$, apparent S\'{e}rsic minor-to-major axis ratio $(b/a)_{\rm obs} = \texttt{sersic\_ba}$, S\'{e}rsic index $n_{\star} = \texttt{sersic\_n}$, and half-light radius along the major axis $r_{\rm eff} = \texttt{sersic\_th50}$.

\section{Methods}
\label{sec:Methods}

Despite the different theoretical background and observational signals, our approach is similar to \cite{Desmond_2018, f(R)_ruled_out}; we forward model the offset, $\bm{r}_{\bullet}$, for the galaxies in our samples and, in conjunction with an empirical noise model describing astrophysical contributions to $\bm{r}_{\bullet}$, derive a likelihood function for the observed offsets for a given $\Delta G / G_{\rm N}$ and $r_{\rm V}$. For a fixed $r_{\rm V}$, we then constrain $\Delta G / G_{\rm N}$ by Markov Chain Monte Carlo (MCMC). As detailed in \Cref{sec:Modelling the Galileon field}, we take $r_{\rm V}$ to be a universal free parameter instead of using a different Vainshtein radius for each galaxy. This will make our constraint on $\Delta G / G_{\rm N}$ at a given $r_{\rm C}$ conservative as we will systematically under-estimate the magnitude of the Galileon field.

We derive the offset expected in Galileon gravity in \Cref{sec:Offset expected from a fifth force} and the gravitational field required to determine this in \Cref{sec:Modelling the gravitational field,sec:Modelling the Galileon field}. In \Cref{sec:Calculating the offset} we convert this to a predicted offset for each galaxy. Using Monte Carlo sampling, we obtain a distribution of offsets, which is modelled as in \Cref{sec:Gaussian mixture model}. We utilise one of the empirical noise models outlined in \Cref{sec:Noise model} to calculate the likelihood function in \Cref{sec:Likelihood Model}. The parameters which are fixed in this section are summarised in \autoref{tab:fixed_parameter_summary}.

\begin{table*}
    \caption{The fixed parameters used to convert dynamical information from the NSA to a predicted offset as described in the text. Above the horizontal line we give the parameters used in the fiducial analysis, and the remainder are used in \Cref{sec:Systematics - Density profile} to test for systematics. In the final column we give the value chosen for each parameter, although we show in \Cref{sec:Discussion} that our results are unchanged for reasonable alternative values.}
    \label{tab:fixed_parameter_summary}
    \centering
    \begin{tabular}{l|l|l}
    \textbf{Parameter} & \textbf{Description} & \textbf{Value} \\
    \hline
    \texttt{ba\_min} & The minimum allowed minor-to-major axis ratio. & 0.15 \\
    $\sigma_{\rm M}$ & Scatter (dex) in $M_{\star}-M_{\rm gas}$ relation (\autoref{eq:Mstar-Mgas relation}). & 0.3 \\
    $\sigma_{\rm R}$ & Scatter (dex) in $R_{\rm eff,gas} - R_{\rm eff}$
    relation (\autoref{eq:Rgas-Rgal relation}). & 0.25 \\
    $\sigma_{\rm D}$ & Additional scatter (dex) in dynamical surface density. & 0.5 \\
    $n_{\rm gas}$ & S\'{e}rsic index for gas component & 1 \\
    \texttt{nsim} & The number of constrained simulations used to reconstruct the gravitational field. & 106 \\
    $N$ & The number of grid points per side length used to reconstruct the gravitational field. & 512 \\
    $\ell$ & The size of the box used to create artificial long wavelength modes for the & 6\\
    & gravitational field, in units of the box length of the constrained simulations. & \\
    \texttt{N\_MC} & The number of Monte Carlo runs to get the distribution of offsets for the template signal. & 500,000 \\
    \hline
    \texttt{N\_AM} & The number of abundance matching realisations. & 200 \\
    $n$ & Slope of halo density profile in the central regions of the halos. & $0 \leq n < 1$ \\
    \end{tabular}
\end{table*}

\subsection{Offset expected from a fifth force}
\label{sec:Offset expected from a fifth force}

In this section we briefly summarise the derivation of \cite{Hui_2012} for the predicted offset of a BH from the galactic centre in Galileon gravity.

In the rest frame of the galaxy, the black hole equation of motion is
\begin{equation} \label{eq:BH EOM}
	\ddot{r} = - \frac{G_N M \left(<r\right)}{r^2} + a_{\rm BH},
\end{equation}
where $M \left(<r\right)$ is the mass enclosed at a distance $r$ from the galaxy's centre. Since the galaxy receives an additional acceleration from the fifth force, the restoring force on the BH must be the same for equilibrium. Using \autoref{eq:fifth force for Galileon}, this is
\begin{equation} \label{eq:BH acceleration}
	\bm{a}_{\rm BH} = - \alpha \nabla \varphi_{\rm ext} = - \left(\frac{\Delta G}{G_N} \right) \nabla \Phi_{\rm lss} = \left(\frac{\Delta G}{G_N} \right) \bm{g}_{\rm lss},
\end{equation}
where $\Phi_{\rm lss}$ and $\bm{g}_{\rm lss}$ are the gravitational potential and acceleration sourced by large scale structure respectively. The Galileon force is therefore proportional to the regular gravitational force. Note that this is only true in the linear regime of the Galileon, where it satisfies a Poisson equation identical to the gravitational potential up to a normalisation factor of $2\alpha$.

Since the mean predicted offset is $\mathcal{O}(10{\rm \, pc})$ for $\Delta G/ G_{\rm N} = 1$, we are only interested in the very central regions of our galaxies. We therefore assume a constant density, $\rho_0$, giving an enclosed mass $M (<r) = 4\pi\rho_0 r^3/3$. At equilibrium $\ddot{r}=0$, so the offset between the BH and the centre of the galaxy, $r_{\bullet}$, is
\begin{equation}\label{eq:cored_offset}
	r_{\bullet} = \frac{3}{4\pi} \frac{\Delta G}{G_N}   \frac{\left| \bm{g}_{\rm lss} \right|}{G_{\rm N}\rho_{0}}.
\end{equation}

We define the right ascension (J2000; RA; $\alpha$) and declination (Dec; $\delta$) directions on the plane of the sky and decompose the offset into these components. Since the observations are two dimensional, from now on we will use $\bm{r}_{\bullet}$ to refer to $r_{\bullet, \alpha} \hat{\alpha} + r_{\bullet, \delta} \hat{\delta}$ and we define the angular offset
\begin{equation} \label{eq:Angular offset}
    \bm{\theta}_{\bullet} \equiv \frac{\bm{r}_{\bullet}}{d_{\rm A}} = \theta_{\bullet, \alpha} \hat{\alpha} + \theta_{\bullet, \delta} \hat{\delta},
\end{equation}
for angular diameter distance $d_{\rm A}$.

\subsection{Modelling the gravitational field}
\label{sec:Modelling the gravitational field}

In this section we describe how we model the gravitational field using $\Lambda$CDM N-body simulations and the matter power spectrum for a $\Lambda$CDM cosmology. Although for given initial conditions the presence of a Galileon increases power on large scales \cite{Barreira_2015}, our use of concordance cosmological parameters reproduces the observed matter power spectrum \cite{Planck_I_2018}, justifying our use of these as an approximation to the density field for a GR+Galileon cosmology. Nevertheless, if the power was enhanced on large scales then we would be systematically underestimating the magnitude of the gravitational (and hence Galileon) field and therefore overestimating the strength of the coupling. This makes our constraints on $\Delta G / G_{\rm N}$ conservative. This is further justified post-hoc by our tight constraint on fifth-force strength, which ensures that differences between the $\Lambda$CDM and `true' density fields cannot be large.

\subsubsection{Constrained simulations of the local volume}
\label{sec:Constrained simulations}

We introduce \texttt{CSiBORG} (Constrained Simulations in BORG), a suite of constrained N-body simulations of the local universe. These are based on the inference of the \texttt{BORG} (Bayesian Origin Reconstruction in Galaxies) algorithm \cite{BORG_1,BORG_2,BORG_3,BORG_4,Lavaux}, which reconstructs the initial conditions (ICs) of the local dark matter density field by applying a Bayesian forward model for the number densities of observed galaxies in voxels, marginalising over galaxy bias parameters. The full \texttt{CSiBORG} suite takes $\sim$100 sets of $z=69$ ICs from the posterior of the particle-mesh \texttt{BORG} reconstruction of the 2M++ volume \cite{2Mpp,BORG-PM} separated by several autocorrelation lengths. These cover a box length of $677.77 {\rm \, Mpc}/h$ with $256^3$ voxels, yielding a resolution of $2.65 {\rm \, Mpc}/h$. Within a smaller sphere of radius $155 {\rm \, Mpc}/h$ centred on the Milky Way we augment the ICs with white noise to a resolution of $2048^3$, giving a particle mass of $2.2 \times 10^8 M_\odot$. We then use each set of ICs to run a DM-only \texttt{RAMSES} \cite{ramses} simulation to $z=0$, refining only in the higher-resolution central sphere (although keeping the larger cube to include longer-wavelength modes). This produces $\sim$100 N-body realisations of the local DM structure from which we can calculate both the magnitude and direction of the $\bm{g}$ field at any nearby point in space. By sampling the realisations we marginalise over both the uncertainties in the constraints on the ICs derived from 2M++ and the unconstrained smaller-scale modes, and hence over the local DM density field itself. \texttt{BORG} and \texttt{CSiBORG} use the cosmology $T_\text{CMB}=2.728$ K, $\Omega_\text{m} = 0.307$, $\Omega_\Lambda = 0.693$, $\Omega_\text{b} = 0.04825$, $H_0 = 70.5$ km s$^{-1}$ Mpc$^{-1}$, $\sigma_8 = 0.8288$, $n=0.9611$. We anticipate the \texttt{CSiBORG} suite to be useful for a range of applications, and we will make it available upon request.

We model the density fields produced in the simulations, $\Delta (\bm{x})$, by applying a cloud-in-cell algorithm to the dark matter particles and solve Poisson's equation 
\begin{equation}
\label{eq:Poisson}
	\Phi \left( \bm{k} \right) = - \frac{4 \pi G_{\rm N} \bar{\rho}}{k^2} \Delta \left( \bm{k} \right),
\end{equation}
on a grid with $N=512$ grid points per side, where $\Delta (\bm{k})$ is the Fourier transform of $\Delta (\bm{x})$. We have checked that using a coarser resolution ($N =256$) does not affect our results.

\subsubsection{Adding larger scale modes}
\label{sec:Adding larger scale modes}

Due to the finite size of the box, we do not have information about the gravitational field all the way down to $k=0$; we can only construct modes with $k \geq \pi / L$ for box length $L$. To fully reconstruct the gravitational field, we must therefore add in long wavelength modes. Unlike the modes captured by the constrained simulation, we have no constraints on the direction of these, so each mode is added as a noise term with a random orientation.

We start by generating a continuous matter power spectrum, $P(k)$, using \texttt{CLASS} \cite{Blas_2011}, assuming a $\Lambda$CDM cosmology with the same parameters as in \Cref{sec:Constrained simulations}. We then construct a grid of size $L^\prime = \ell L$ for $\ell > 1$ with $N^\prime$ grid points, such that the maximum $k$ obeys $k_{\rm max} \geq \pi / L$. This ensures that the modes added here begin where those in the simulation boxes end. 

A Gaussian random field, $\psi (\bm{k})$, is generated on the grid, with the condition $\psi (\bm{k}) = \psi^{\ast} (-\bm{k})$ to ensure the density contrast is real. To determine this, we must adapt the continuous power spectrum for the discrete case \cite{Pen_1997,Bertschinger_2001}; we must account for the normalisation in our Fourier convention, the change in measure and the units of the power spectrum. After doing this, we can obtain the potential by solving \autoref{eq:Poisson}.

We filter out all modes which overlap between the large and small boxes to prevent double counting and inverse Fourier transform to obtain $\bm{g} \left( \bm{r} \right)$ from these large scale modes. These are added to the field obtained from the constrained simulation. Since the magnitude and direction of the modes added in the above procedure are not constrained, we must marginalise over the direction and magnitude by incorporating this into our Monte Carlo sampling. The addition of long wavelength modes increases the root mean square of $|\bm{g}|$ by 5\% and scatters each Cartesian component by $\sim 17\%$ compared to their uncorrected values.

This process required a further parameter in our inference: the size of the large box, $L^\prime = \ell L$. In \Cref{app:Box size for long wavelength modes} we show that using $\ell=6$ is appropriate given the uncertainty on $\bm{g}$ from the constrained simulations and investigate the choice of $\ell$ further in \Cref{sec:Systematics: Galileon field}.

\subsection{Modelling the Galileon field}
\label{sec:Modelling the Galileon field}

In \autoref{eq:BH acceleration} we assume that the fifth force is proportional to the gravitational field sourced by large scale structure, while on small scales the Galileon is screened. As a model for this, we take the field calculated in \Cref{sec:Modelling the gravitational field} and apply a low-pass filter, such that we remove all $k$ modes with $|\bm{k}| > k_{\rm V}$, corresponding to a Vainshtein radius for large scale structure of
\begin{equation}
    r_{\rm V} \equiv \frac{2 \pi}{k_{\rm V}}.
\end{equation}
We choose a constant $r_{\rm V}$ for all galaxies, instead of filtering at the scale of each galaxy's Vainshtein radius, $r_{\rm V}^{\left(g\right)}$, as a function of its mass and $r_{\rm C}$ (\autoref{eq:Vainshtein radius definition}). This essentially corresponds to an average over all galaxies, and simplifies the analysis because we do not have to apply a different filter for each galaxy or re-filter each time we change $\Delta G / G_{\rm N}$ in our inference which would require us to re-derive the likelihood by running the Monte Carlo sampling again.

To convert our constraint on $\Delta G / G_{\rm N}$ as a function of $r_{\rm V}$ to one on $\alpha$ as a function of $r_{\rm C}$, we require $r_{\rm C}$ as a function of $r_{\rm V}$ and $\alpha$. To determine this in a cosmological context, we consider the mass enclosed within radius $r$ due to cosmological perturbations:
\begin{equation}
    M \left( r \right) = \bar{\rho} \int_{|\bm{x}|<r} \Delta \left(\bm{x} \right) {\rm d}^3 x.
\end{equation}
We only consider the contribution from perturbations and not the background because the gravitational effects of the latter are encoded in the evolution of the Hubble parameter \cite{Lue}. The mean square value is determined by the matter power spectrum,
\begin{equation}
    \label{eq:Cosmological mass enclosed}
    \left< M^2 \left( r \right) \right> = \left( 4 \pi \bar{\rho} \right)^2 \int \frac{{\rm d}^3 k}{\left( 2 \pi \right)^3} \frac{P \left( k \right)}{k^4} \left( \sin \left( kr \right) - kr \cos \left( kr \right) \right)^2,
\end{equation}
where we use \texttt{CLASS} to compute the non-linear $P(k)$ for a $\Lambda$CDM cosmology. Using \autoref{eq:Vainshtein radius definition}, we can thus determine $r_{\rm C}$ as 
\begin{equation}
    \label{eq:rv to rc conversion}
    r_{\rm C} \simeq \frac{1}{3} \left( \frac{3 r_{\rm V}^3}{4\alpha G \left< M^2 \left( r_{\rm V} \right) \right>^{\frac{1}{2}}} \right)^{\frac{1}{2}} ,
\end{equation}
where the arbitrary factor of $1/3$ is included so that $r_{\rm V} = 10 {\rm \, Mpc}$ corresponds to $r_{\rm C} \sim 6 {\rm \, Gpc}$, as found in simulations \cite{Chan_2009}. We run our inference for a range of $r_{\rm V} > 1 {\rm \, Mpc}$, with $r_{\rm V}=100{\rm \, Mpc}$ as our fiducial case.

Using the power spectrum to obtain the covariance of $\Delta (\bm{k})$, we can find the expectation value of the square of the gravitational field from large scale structure\footnote{This definition is identical to \autoref{eq:g_cts correction definition} and so $\delta g_{\rm cts}^2$ is given by \autoref{eq:g_cts correction result}.}, $\delta g_{\rm cts}^2$, by squaring \autoref{eq:Poisson} and only keeping modes $|\bm{k}| < k_{\rm V}$. Assuming $P(k) \propto k$ for $k < k_{\rm eq}$ where $L_{\rm eq} = 2 \pi / k_{\rm eq} \approx 450 {\rm \, Mpc}$, for two values of $r_{\rm V} < L_{\rm eq}$, $r_1$ and $r_2$, we find the fractional difference in the field is
\begin{equation}
    \frac{ \left( \delta g_{\rm cts} \left( r_1 \right) ^2 \right)^{\frac{1}{2}} - \left( \delta g_{\rm cts}  \left( r_2 \right)^2 \right)^{\frac{1}{2}}}{\left( \delta g_{\rm cts}  \left( r_1 \right)^2 \right)^{\frac{1}{2}}} = 1 -\sqrt{ \frac{1 - \frac{1}{2} \left( \frac{r_{2}}{L_{\rm eq}} \right)^{2}}{1 - \frac{1}{2} \left( \frac{r_{1}}{L_{\rm eq}} \right)^{2}} }.
\end{equation}
This fractional difference is only 1 per cent between $r_{\rm V} = 1 {\rm \, Mpc}$ and $r_{\rm V} = 100 {\rm \, Mpc}$ or 5 per cent between $r_{\rm V} = 1 {\rm \, Mpc}$ and $r_{\rm V} = 200 {\rm \, Mpc}$. This implies that the calculated field is relatively insensitive to the choice of $r_{\rm V}$ provided $r_{\rm V} < L_{\rm eq}$.

In reality the separation between screened and unscreened modes will be more gradual than this step-function filter. However, this insensitivity to $r_{\rm V}$ suggests that a smoother filter will not dramatically change our results. 

This model neglects the impact of the non-linear regime and requires the Galileon to be linear. This, combined with the insensitivity of the fifth force field to $r_{\rm V} < L_{\rm eq}$ means our constraints are valid for $ 10 {\rm \, Mpc} \lesssim r_{\rm V} \lesssim 450 {\rm \, Mpc}$.

\subsection{Calculating the offset}
\label{sec:Calculating the offset}

In order to the calculate the magnitude of the offset, $r_{\bullet}$, we need to know the total enclosed mass within separation $r_{\bullet}$, $M(<r_{\bullet})$. In the absence of central kinematic data in our observational datasets, we must either attempt to fit a density profile to the galaxies using empirical methods such as abundance matching (AM) or employ empirical scalings between mass and light at the centre of the galaxies. We will find that $r_{\bullet} \ll$ size of the galaxy, so that the latter is more reliable as the former requires an integral over the galaxy's full luminosity profile. We therefore use this method.

Assuming a cored density profile, we wish to find the (constant) central density, $\rho_{0}$, using the information obtained from the NSA. To do this we must determine the major and minor axis lengths from the observed minor-to-major axis ratio, $(b/a)_{\rm obs}$, effective radius, $r_{\rm eff}$, and redshift, $z_{\rm dist}$. We also use the measured stellar mass, $M_{\star}$, and intensity profile to determine the central surface density, utilising observed correlations to estimate the contributions from dark matter and gas. Combining these two results gives us $\rho_{0}$. 

Using the observed minor-to-major axis ratio $(b/a)_{\rm obs}$ from the NSA, we assign a random inclination, $i$, to the galaxy and estimate the true axis ratio to be
\begin{equation}
	\left( \frac{b}{a} \right)^2 = 1 - \frac{1 - (b/a)_{\rm obs}^2}{\sin^2 i},
\end{equation}
with the condition that $(b/a) \geq \texttt{ba\_min}$, where $\texttt{ba\_min} = 0.15$ since this is the lowest axis ratio recorded in the NSA.

We use \texttt{zdist} from the NSA catalogue to determine the angular diameter distance $d_A$ to each galaxy and calculate the major-axis length as $R_{\rm eff}^{\rm maj} \equiv d_A r_{\rm eff}$. This is related to circularised, $R_{\rm eff}$, and minor-axis, $R_{\rm eff}^{\rm min}$, half-light radii as
\begin{equation}
	R_{\rm eff} = \left( \frac{b}{a} \right)^{\frac{1}{2}} R_{\rm eff}^{\rm maj}, \qquad R_{\rm eff}^{\rm min} = \left( \frac{b}{a} \right) R_{\rm eff}^{\rm maj}.
\end{equation}

We now deproject the S\'{e}rsic profile to find the stellar surface density \cite{Prugniel_1997,Ciotti_1999}
\begin{equation}
\label{eq:Sersic central density}
	\Sigma_{\star} = \frac{M_{\star}}{2 \pi b_{n_{\star}}^{-2n_{\star}} \Gamma \left( 2 n_{\star} \right) R_{\rm eff}^2},
\end{equation}
where $b_{n_{\star}} \equiv 2n_{\star} - 1/3 + 0.009876 / n_{\star}$. Using the reverse of the method from \cite{Desmond_2018}, we can estimate the gas mass from the stellar mass using \cite{Dutton_2011}
\begin{equation}
\label{eq:Mstar-Mgas relation}
	\log_{10} \left( \frac{M_{\star}}{M_\odot} \right) = 1.89 \log_{10} \left( \frac{M_{\rm gas}}{M_\odot} \right) - 8.12,
\end{equation}
with scatter $\sigma_{\rm M} = 0.3 {\rm \, dex}$. Assuming an exponential disk with effective radius given by
\begin{equation}
\label{eq:Rgas-Rgal relation}
	\log_{10} \left( \frac{R_{\rm eff, gas}}{{\rm kpc}} \right) = \log_{10} \left( \frac{0.92 R_{\rm eff}}{{\rm kpc}} \right),
\end{equation}
with scatter $\sigma_{\rm R} = 0.25 {\rm \, dex}$, we can calculate the central gas surface density, $\Sigma_{\rm gas}$, using \autoref{eq:Sersic central density} with the appropriate mass and radius and with $n_{\rm gas}=1$ instead of $n_{\star}$. We now have the central baryonic surface density
\begin{equation}
	\Sigma_B = \Sigma_{\star} + \Sigma_{\rm gas}.
\end{equation}
To convert this to the central dynamical surface density, $\Sigma_{\rm D}$, we use the empirical relation  \cite{lelli2014inner,Lelli_2016,Milgrom_2016}
\begin{equation}
	\Sigma_{\rm D} = \Sigma_{\rm M} S \left( \frac{\Sigma_{\rm B}}{\Sigma_{\rm M}} \right),
\end{equation}
where
\begin{equation}
	S\left( y \right) = \frac{y}{2} + y^{\frac{1}{2}} \left( 1 + \frac{y}{4} \right)^{\frac{1}{2}} + 2 \sinh^{-1} \left( \frac{ y^{\frac{1}{2}}}{2} \right),
\end{equation}
with $\Sigma_{\rm M} = 1.37 \times 10^8 M_\odot {\rm kpc^{-2}}$. Despite the already large scatter due to uncertainties on the input quantities, one may expect the scaling relations to provide good fits only to a subset of the galaxy population. Therefore, to ensure our constraints on $\Delta G / G_{\rm N}$ are conservative, we impose an additional scatter of $\sigma_{\rm D} = 0.5 {\rm \, dex}$. For scale height $h$, which we assume is equal to $R_{\rm eff}^{\rm min}$, $\Sigma_{\rm D}$ is related to the central density as
\begin{equation}
	\Sigma_{\rm D} \equiv 2 h \rho_{0},
\end{equation}
which we can substitute into \autoref{eq:cored_offset} to determine $r_{\bullet}$.

\subsection{Gaussian mixture model}
\label{sec:Gaussian mixture model}

From \autoref{eq:cored_offset}, we see that the offset is proportional to $\Delta G / G_{\rm N}$. We therefore construct a template signal with $\Delta G / G_{\rm N} = 1$ containing $N_{\rm MC}$ realisations of our probabilistic model.

For a given $r_{\rm V}$, we must convert the $N_{\rm MC}$ samples of predicted $\theta_{\bullet, \alpha}$ and $\theta_{\bullet, \delta}$ for each galaxy into a distribution. Following \cite{f(R)_ruled_out}, we model the samples as a Gaussian mixture model (GMM) \cite{scikit-learn} where the likelihood function for some galaxy $g$ is
\begin{equation}
	\mathcal{L}_{\rm g} \left( \theta_{\bullet, \alpha}  | \Delta G, r_{\rm V} \right) = \sum_i \frac{w^{\left( i \right)}_{\rm g, \alpha}}{\sqrt{2 \pi}\sigma^{\left( i \right)}_{\rm g, \alpha}} \exp \left[ - \frac{\left( \theta_{\bullet, \alpha} - \mu^{\left( i \right)}_{\rm g, \alpha}\right)^2}{2 {\sigma^{\left( i \right)}_{\rm g, \alpha}}^2} \right],
\end{equation}
where
\begin{equation}
	\sum_i w_{\rm g, \alpha}^{\left( i \right)} = 1, \quad  w_{\rm g, \alpha}^{\left( i \right)} \geq 0,
\end{equation}
and $\{ w^{\left( i \right)}, \sigma^{\left( i \right)},\mu^{\left( i \right)} \}$, the weights, standard deviations and means of the Gaussians, are implicit functions of $r_{\rm V}$. There is an analogous definition for the declination component. The sum runs over the number of Gaussian components. The number of components is chosen to minimise the Bayesian Information Criterion (BIC)
\begin{equation}
\label{eq:BIC}
	{\rm BIC} = \mathcal{K} \log \mathcal{N} - 2 \hat{\mathcal{L}},
\end{equation}
for $\mathcal{K}$ model parameters, $\mathcal{N}=N_{\rm MC}$ data points, and maximum likelihood estimate $\hat{\mathcal{L}}$. We find an independent set of Gaussians for each galaxy and component. For a different value of $\Delta G /G_{\rm N}$, we must transform the means and widths of the Gaussians in the GMM
\begin{equation} \label{eq:Scaling GMM}
	\tilde{\mu}^{\left( i \right)}_{\rm g, \alpha} = \left( \frac{\Delta G}{G_{\rm N}} \right) \mu^{\left( i \right)}_{\rm g, \alpha}, \quad \tilde{\sigma}^{\left( i \right)}_{\rm g, \alpha} = \left( \frac{\Delta G}{G_{\rm N}} \right) \sigma^{\left( i \right)}_{\rm g, \alpha}.
\end{equation}

 Treating the orthogonal RA and Dec components as independent, the overall likelihood $\mathcal{L}_{\rm g} ( \theta_{\bullet, \alpha}, \theta_{\bullet, \delta} )$ for a test galaxy $g$ to have $\bm{\theta}_{\bullet}$ components $\theta_{\bullet, \alpha}$ and $\theta_{\bullet, \delta}$ is 
\begin{equation}
    \begin{split}
	\mathcal{L}_{\rm g} & \left(\theta_{\bullet, \alpha}, \theta_{\bullet, \delta} | \Delta G, r_{\rm V} \right) \\ 
	&= \mathcal{L}_{\rm g}\left(\theta_{\bullet, \alpha}| \Delta G, r_{\rm V} \right) \mathcal{L}_{\rm g}\left(\theta_{\bullet, \delta}| \Delta G, r_{\rm V} \right).
	\end{split}
\end{equation}

\subsection{Modelling the noise}
\label{sec:Noise model}

Galileons are not the only type of physics that can lead to BH--galaxy offsets, requiring us to develop a model for astrophysical noise. Some examples of other relevant physics include three-body interactions between BHs following two successive mergers \citep{Hut_1992,Xu_1994}; subhalo accretion, which can cause offsets of tens of parsecs by transferring energy to the BH by dynamical friction \citep{boldrini2020subhalo}; and gravitational wave emission from a BH binary at the centre of the galaxy, which, by linear momentum conservation, causes the centre of mass to recoil \citep{Peres_1962,Bekenstein_1973}. A population of offset and wandering BHs \citep{Volonteri_2003} is therefore expected even in the absence of a fifth force.

We must also consider the noise due to observational errors. Central BHs may appear to be offset due to misassociation, extended sources, double or lensed quasars, statistical outliers due to an extended tail \citep{Mingard_2018}, the presence of a jet \citep{Kovalev_2017,Petrov_2017b} or dust lanes.

Ideally we would construct a model for the observed offsets in the absence of a fifth force by using cosmological hydrodynamical simulations. Unfortunately the majority of these simulations do not include a prescription for dynamical friction on BHs and instead pin the BHs at the centre of the galaxy \cite{Taylor_2014,Schaye_2015,Sijacki_2015}. Those that do allow the BH to move over-predict the fraction of offset BHs compared to observations \cite{Bartlett_2020}. We therefore must construct an empirical noise model based on the global distribution of offsets in the various datasets.

We consider three different noise models: a Gaussian distribution, the sum of a Gaussian and Laplace distribution, and an Edgeworth expansion. We outline these below, before describing how we discriminate between them. In the following, we define the observed offsets to be $\theta_{\bullet, \alpha, {\rm obs}}$ and $\theta_{\bullet, \delta, {\rm obs}}$, and the true offsets -- which are to be compared to the fifth force prediction -- are $\theta_{\bullet, \alpha}$ and $\theta_{\bullet, \delta}$.

\subsubsection{Gaussian Noise Model}

In this model we assume that the observed value is Gaussian distributed about the predicted value due to a fifth force, such that the observed value has some uncertainty $\sigma_{\rm obs}$. We assume this is equal for both the RA and Dec components. We therefore have
\begin{equation}
    \mathcal{L}_{\rm g}\left(\theta_{\bullet, \alpha, {\rm obs}}|\theta_{\bullet, \alpha},\bm{\Omega} \right) = \frac{1}{\sqrt{2 \pi} \sigma_{\rm obs}} \exp \left( - \frac{\left(\theta_{\bullet, \alpha, {\rm obs}} - \theta_{\bullet, \alpha} \right)^2}{2 \sigma_{\rm obs}^2}\right),
\end{equation}
with $\bm{\Omega} = \{ \sigma_{\rm obs} \}$.

\subsubsection{Gaussian plus Laplace distribution}

It is known that the distribution of observed BH--galaxy offsets is non-Gaussian \cite{Bartlett_2020}, so we should also consider non-Gaussian noise models. Inspired by \cite{Skipper_2018,Bartlett_2020}, our first non-Gaussian model is the sum of a Gaussian and Laplace distribution,
\begin{equation}
    \begin{split}
        \mathcal{L}_{\rm g}\left(\theta_{\bullet, \alpha, {\rm obs}}|\theta_{\bullet, \alpha},\bm{\Omega} \right) &= \frac{f}{\sqrt{2 \pi} \sigma_{\rm obs}} \exp \left( - \frac{\left(\theta_{\bullet, \alpha, {\rm obs}} - \theta_{\bullet, \alpha} \right)^2}{2 \sigma_{\rm obs}^2}\right) \\
	&+ \frac{1-f}{2 \nu} \exp \left( - \frac{\left| \theta_{\bullet, \alpha, {\rm obs}} - \theta_{\bullet, \alpha} \right|}{\nu} \right),
    \end{split}
\end{equation}
with $\bm{\Omega} = \{ \sigma_{\rm obs}, \nu, f \}$. As with the Gaussian model, $\sigma_{\rm obs}$ describes the uncertainty in the observed value. The Laplace term dominates in the tails of the distribution, hence $\nu$ tells us about the scale to which offset BHs extend, while $f$ is the fraction of the probability in the Gaussian component.

\subsubsection{Edgeworth Expansion}

We now consider another way to incorporate non-Gaussianity, namely through the  Edgeworth expansion~\cite{Blinnikov_1998}
\begin{equation} \label{eq:Edgeworth expansion}
    \begin{split}
	\mathcal{L}_{\rm g} \left(\theta_{\bullet, \alpha, {\rm obs}}|\theta_{\bullet, \alpha},\bm{\Omega} \right) = &\frac{1}{\sqrt{2 \pi} \sigma_{\rm obs}} \exp \left( - \frac{\left(\theta_{\bullet, \alpha, {\rm obs}} - \theta_{\bullet, \alpha} \right)^2}{2 \sigma_{\rm obs}^2}\right) \\
	& \times \sum_{n = 0}^{F} \alpha_n  {\rm H}_n \left(\frac{  \theta_{\bullet, \alpha, {\rm obs}} -  \theta_{\bullet, \alpha} }{\sigma_{\rm obs} \sqrt{2}} \right),
    \end{split}
\end{equation}
where ${\rm H}_n (x)$ are Hermite Polynomials. The parameters $\{\alpha_n\}$ are related, since the probability density must be non-negative. We express $\{\alpha_n\}$ in terms of cumulants and discuss these constraints in \Cref{app:Coefficients in the Edgeworth expansion}. $\bm{\Omega}=\{\sigma_{\rm obs},\alpha_3, \alpha_4, \ldots\}$ in this case, and we consider $3 \leq F \leq 8$.

\subsubsection{Physical offsets}
\label{sec:Physical Offsets}

So far the assumed noise models only contain contributions due to observational effects, so that all noise parameters are angular quantities. This corresponds to the assumption that any noise contribution with a fixed physical scale is subdominant. To account for this possibility, for each noise model we consider the corresponding model with
\begin{equation}
    \sigma_{\rm obs} \to \sqrt{\sigma_{\rm obs}^2 + \left( \frac{\sigma_{\rm int}}{d_{\rm A}} \right)^2},
\end{equation}
which assumes that BHs are isotropically, Gaussian distributed with width $\sigma_{\rm int}$. This adds one more parameter to the noise model and means that each galaxy has a different width of its noise Gaussian.

\subsubsection{Choosing the model}

To decide which noise model to use, for each dataset we find the maximum likelihood estimates, $\hat{\mathcal{L}} \equiv \max \mathcal{L}\left(d|\Delta G, r_{\rm V}, \bm{\Omega}\right)$, for each model. We then choose the model that minimises the BIC (\autoref{eq:BIC}) with $\mathcal{N} = 2 \mathcal{N}_{\rm gal}$ for $\mathcal{N}_{\rm gal}$ galaxies in the dataset. We find that in all cases the sum of the Gaussian and Laplace distribution without a physical offset best describes the data, so this is the noise model we use below. In \autoref{fig:offset_data} we see that the observed angular offsets are independent of redshift, suggesting that these are dominated by observational effects. It is therefore unsurprising that the addition of an intrinsic offset is not required.

\subsection{Likelihood Model}
\label{sec:Likelihood Model}

Now we have a distribution for obtaining an observed value given a predicted one, $\mathcal{L}_{\rm g}\left(\theta_{\bullet, \alpha, {\rm obs}}|\theta_{\bullet, \alpha}, \bm{\Omega} \right)$, after accounting for the noise contribution to the signal. The resulting likelihood for an observed offset $\theta_{\bullet, \alpha, {\rm obs}}$ is
\begin{equation}
    \begin{split}
	\mathcal{L}_{\rm g} & \left(\theta_{\bullet, \alpha, {\rm obs}}| \Delta G, r_{\rm V}, \bm{\Omega}\right) = \\
	& \int \mathcal{L}_{\rm g}\left(\theta_{\bullet, \alpha, {\rm obs}}|\theta_{\bullet, \alpha}, \bm{\Omega} \right) \mathcal{L}_{\rm g}\left( \theta_{\bullet, \alpha}| \Delta G, r_{\rm V}  \right) \mathrm{d}\theta_{\bullet, \alpha}.
	\end{split}
\end{equation}
For example, for the Gaussian noise model, this is 
\begin{equation}
    \begin{split}
    \mathcal{L}_{\rm g} & \left(\theta_{\bullet, \alpha, {\rm obs}}| \Delta G, r_{\rm V}, \bm{\Omega}\right) = \\
    & \sum_i \frac{w^{\left( i \right)}_{\rm g, \alpha}}{\sqrt{2 \pi \left(\sigma_{\rm obs}^2 + \tilde{\sigma}^{\left( i \right)}_{\rm g, \alpha}{}^2 \right)}}   \exp \left( - \frac{\left(\theta_{\bullet, \alpha, {\rm obs}} -  \tilde{\mu}^{\left( i \right)}_{\rm g, \alpha} \right)^2}{2 \left( \sigma_{\rm obs}^2 + {\tilde{\sigma}^{\left( i \right)}_{\rm g, \alpha}}{}^2 \right)}\right) .
    \end{split}
\end{equation}
We treat each galaxy as independent to obtain the likelihood of our dataset $d$ to be
\begin{equation}
    \begin{split}
	\mathcal{L} & \left(d|\Delta G, r_{\rm V}, \bm{\Omega}\right) = \\
	& \prod_{\rm g} \mathcal{L}_{\rm g} \left(\theta_{\bullet, \alpha, {\rm obs}} | \Delta G, r_{\rm V}, \bm{\Omega}\right) \mathcal{L}_{\rm g} \left(\theta_{\bullet, \delta, {\rm obs}} | \Delta G, r_{\rm V}, \bm{\Omega} \right).
	\end{split}
\end{equation}
Finally, given some prior on $\Delta G$ and $r_{\rm V}$, $P\left(\Delta G, r_{\rm V}, \bm{\Omega} \right)$, we use Bayes' theorem to obtain
\begin{equation}
	P\left(\Delta G, r_{\rm V}, \bm{\Omega}| d \right) = \frac{\mathcal{L}\left(d|\Delta G, r_{\rm V}, \bm{\Omega} \right) P\left(\Delta G, r_{\rm V}, \bm{\Omega} \right)}{P\left(d\right)},
\end{equation}	
where $P(d)$ is the constant probability of the data for any $\{ \Delta G, r_{\rm V}, \bm{\Omega} \}$. We are now in a position to derive posteriors on $\Delta G / G_{\rm N}$ and the noise model parameters at fixed $r_{\rm V}$, for which we use the \textsc{emcee} sampler \cite{emcee}. We impose the improper prior $\Delta G \geq 0$, flat in $\Delta G$. The priors for all inferred parameters are given in \autoref{tab:infered_parameter_summary}.

\begin{table}
    \caption{Inferred parameters describing the predicted signal and the empirical noise models. In all cases the Gaussian plus Laplace distribution is the preferred noise model, and the parameters for this are given in the top part of the table. Below the horizontal line we give the parameters for the Edgeworth expansion (\Cref{eq:Edgeworth expansion,eq:Define Edgeworth parameters}) and the models containing intrinsic offsets, which we use in \Cref{sec:Systematics - Noise model}. All priors are uniform in the range given. We fix $r_{\rm V}$ in the inference and allow all other parameters relevant to the chosen noise model to vary. The $(p, q)$ values refer to \autoref{eq:Cauchy-Bunyakovskii}.}
    \label{tab:infered_parameter_summary}
    \centering
    \begin{tabular}{c|c}
    Parameter & Prior/Constraint \\
    \hline
    $ \Delta G / G_{\rm N}$ & $ \geq 0$\\ 
    $\log_{10} \left( r_{\rm V} \ / \ {\rm Mpc} \right)$ & $\geq 0.5 $\\
    $\sigma_{\rm obs}$ & $ > 0$\\
    $\nu$ & $ > 0$\\
    $f$ & $[0, 1]$\\
    \hline
    $\sigma_{\rm int}$ & $ \geq 0$\\
    $\gamma$ & - \\
    $\tau$ & $p=1$, $q=2$\\
    $\eta$ & - \\
    $\zeta$ & $p=2$, $q=3$ \\
    $\xi$ & - \\
    $\iota$ & $p<q$, $q=4$\\
    \end{tabular}
\end{table}

\section{Results}
\label{sec:Results}

\begin{figure*}
	\centering
	\includegraphics[width=\textwidth]{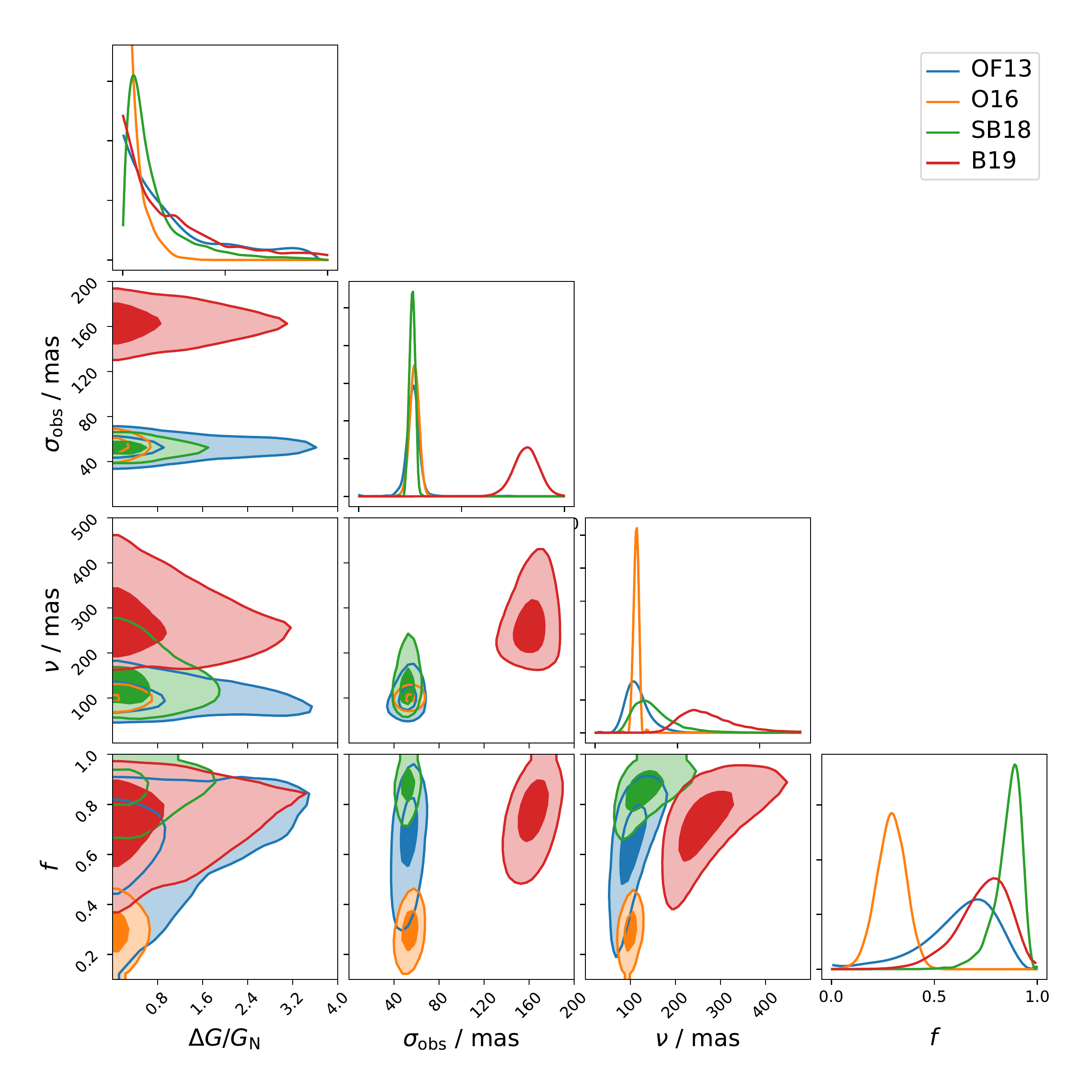}
	\caption{\label{fig:results_corner}Corner plot of the constraints on the strength of the coupling to the Galileon field, $\Delta G / G_{\rm N}$, and the noise parameters at $r_{\rm V} = 100 {\rm \, Mpc}$. The contours show the $1$ and $2\sigma$ confidence intervals. Each dataset is consistent with $\Delta G / G_{\rm N} = 0$, with $\Delta G / G_{\rm N} > 0.16$ ruled out at $1\sigma$ confidence when we combine the datasets. Note that while $\Delta G / G_{\rm N}$ is assumed universal, $\sigma_\text{obs}$, $\nu$ and $f$ are sample-specific.}
\end{figure*}

In \autoref{fig:results_corner} we show the corner plot from the inference with $r_{\rm V} = 100 {\rm \, Mpc}$ for each dataset, using the empirical noise model consisting of a Gaussian plus a Laplace distribution. We see that each dataset is consistent with $\Delta G / G_{\rm N} = 0$. Assuming each sample is independent, we multiply the likelihoods and, giving each dataset a different set of noise parameters, find the joint constraint of
\begin{equation}
    \frac{\Delta G}{G_{\rm N}} < 0.16
\end{equation}
at $1\sigma$ confidence for this value of $r_{\rm V}$, or $< 0.36$ at $2 \sigma$ confidence. We find that the constraint is driven by O16 due to its large size. If we did not include the O16 data, our strongest constraint would come from SB18 and would be $\Delta G / G_{\rm N} < 0.65$ at $1\sigma$ confidence.

For the radio samples (OF13, O16 and SB18) we find that $\sigma_{\rm obs} \sim 50 {\rm \, mas}$ and for B19 we find $\sigma_{\rm obs} \sim 150 {\rm \, mas}$, as expected in \Cref{sec:Observational data}. Further, as in \cite{Bartlett_2020}, we find that O16 has a much higher contribution from the non-Gaussian component than the other datasets, as shown by the smaller value of $f$.

\begin{figure}
	\centering
	\includegraphics[width=\columnwidth]{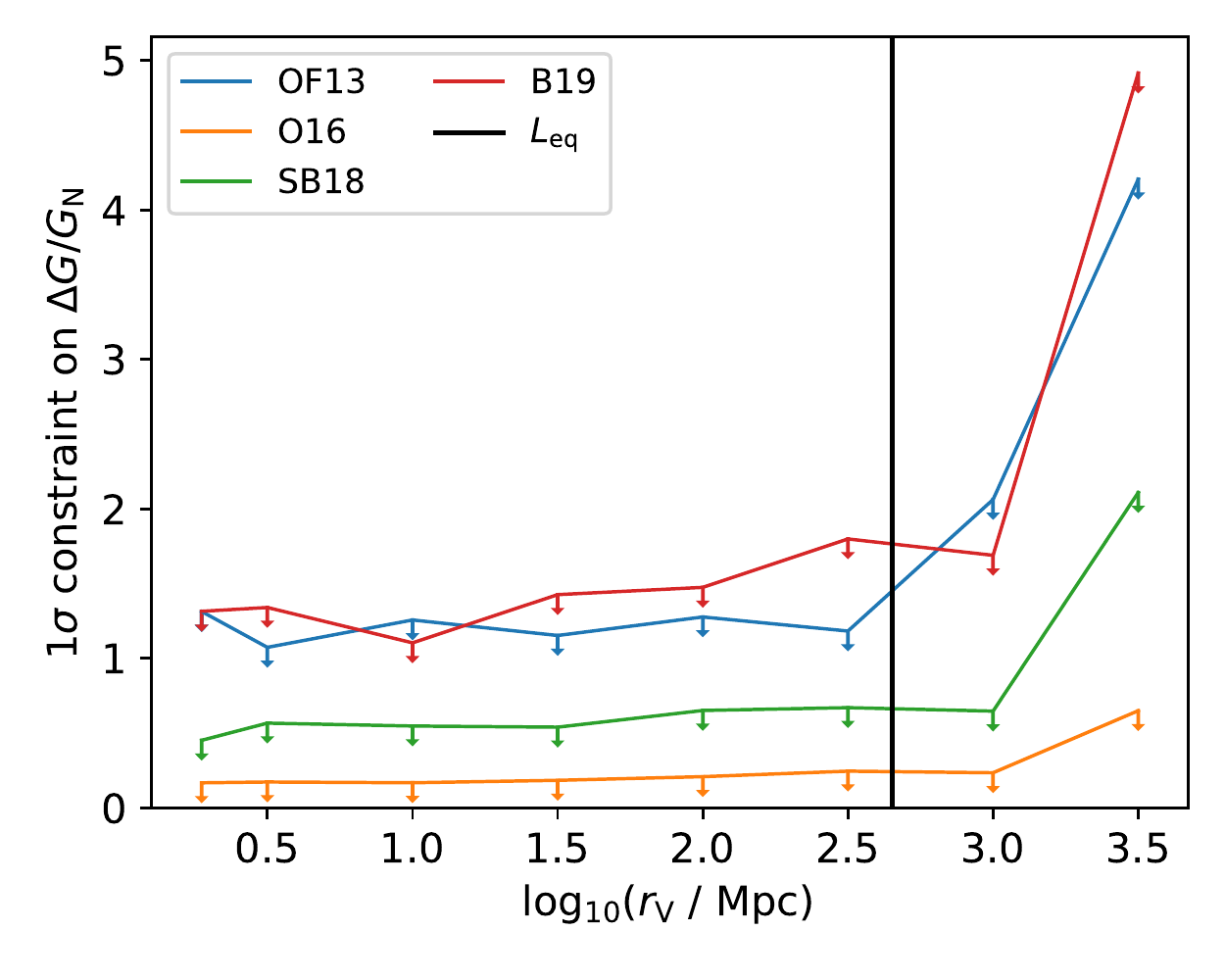}
	\caption{\label{fig:DG vs LV}$1\sigma$ constraint on $\Delta G / G_{\rm N}$ as a function of average Vainshtein radius, $r_{\rm V}$. We expect $r_{\rm V} \sim 10 {\rm \, Mpc}$ for crossover scales $\sim H_0^{-1}$. $r_{\rm V} > L_{\rm eq}$ corresponds to crossover scales much larger than the observable universe (see \autoref{fig:alpha-rc constraint}).}
\end{figure}

We repeat the inference at different values of $r_{\rm V} > 1 {\rm \, Mpc}$ and plot the $1\sigma$ constraint as a function of $r_{\rm V}$ for each dataset in \autoref{fig:DG vs LV}. As anticipated in \Cref{sec:Modelling the Galileon field}, we find that the constraint is relatively independent of $r_{\rm V}$ for $r_{\rm V} < L_{\rm eq}$, where the level of bumpiness for these $r_{\rm V}$ indicates the noise level of this method, due to the finite number of Monte Carlo realisations used to determine the likelihood. To quantify this, we run the end-to-end inference a further 6 times for the OF13 data with $r_{\rm V} = 100 {\rm \, Mpc}$ and find an unbiased sample variance of the $1\sigma$ and $2\sigma$ constraints of 8\% and 7\% respectively. This shows that the number of Monte Carlo realisations is sufficiently large.

The lack of dependence of the constraint on $r_{\rm V}$ means we expect our constraints will not change for a broad transition from screened to unscreened in this regime, as opposed to the step-function we currently use.

For larger values of $r_{\rm V}$ we find that, although the posteriors are still consistent with $\Delta G / G_{\rm N} = 0$, the constraint weakens. The smaller magnitude of the fifth force field at these $r_{\rm V}$ (more modes of the $\bm{g}$ field are excluded) means a given offset requires a larger value of $\Delta G / G_{\rm N}$ (\autoref{eq:cored_offset}), and thus a worse constraint. $r_{\rm V} = L_{\rm eq}$ corresponds to the largest crossover scales in \autoref{fig:alpha-rc constraint}, which are already much larger than $H_0^{-1}$.

\section{Discussion}
\label{sec:Discussion}

\subsection{Systematic uncertainties}

In this section we vary several parameters of the analysis which could contribute systematic error if kept fixed. For computational convenience, throughout this section we use the OF13 data as this is the smallest dataset, although the conclusions apply equally to all the datasets. Hence the constraint for the fiducial case in this section is $\Delta G / G_{\rm N} < 1.28$, as opposed to the full joint constraint of $\Delta G / G_{\rm N} < 0.16$.

\subsubsection{Density profile}
\label{sec:Systematics - Density profile}

Although halos in N-body simulations are typically well fit by NFW profiles, the situation is less clear observationally and in the presence of baryons. While the process of adiabatic contraction steepens the central DM density during the process of galaxy formation \cite{Blumenthal, Gnedin}, subsequent stellar feedback can inject energy into the halo and cause it to expand (e.g. \cite{Pontzen_Governato, DP_CuspCore}). Previously we assumed that this results in an approximately uniform density at the centre of the halo, however this need not be true.

To test whether our constraints are sensitive to the assumed density profile, we now suppose that the inner density can be described by a power law. To remain agnostic as to the slope of the DM density profile over the extent of the galaxy, we parametrise it as
\begin{equation}
    \label{eq:power law density profile}
	\rho\left(r\right) =  
  \begin{cases}
    \rho_{\rm s} \left(\frac{r_{\rm s}}{r}\right)^n, & r \leq r_{\rm s} \\
    \frac{4 \rho_{\rm s}}{r/r_{\rm s} \left( 1 + r/r_{\rm s}\right)^2}, & r > r_{\rm s},
  \end{cases}
\end{equation}
with $n$ a free parameter that would equal 1 for NFW. With most observational evidence favouring profiles somewhat shallower than NFW (e.g.~\cite{Cusp_core,Oh,Salucci}) we take a fiducial value for $n$ of 0.5, although we will check explicitly the result of varying it within the a priori plausible range $0 \leq n < 1$. $r_s$ is the scale radius of the halo, while $\rho_s$ is the density at that radius.

For $n=0.5$, the mean predicted offset is $\sim50{\rm \, pc})$ for $\Delta G/ G_{\rm N} = 1$. This is much less than $r_{\rm s}$ so that the relevant profile is the power law, and the NFW profile is only required to determine $\rho_{\rm s}$. We therefore do not need to consider the case where the restoring force is too small to balance the fifth force, which could lead to the BH being ejected from the galaxy \cite{Sakstein_2017}.

Within the power-law region the enclosed mass is
\begin{equation}
	M \left(<r\right) = \frac{4\pi}{3-n} \rho_{\rm s} r^3 \left(\frac{r_{\rm s}}{r}\right)^n, \quad r < r_{\rm s}.
\end{equation}
Substituting this into \autoref{eq:BH EOM}, the equivalent of \autoref{eq:cored_offset} is
\begin{equation} \label{eq:power_law_offset}
	r_{\bullet} = \left( \frac{\Delta G}{G_N} \right)^\frac{1}{1-n} \left( \frac{3-n}{4\pi} \frac{\left| \bm{g}_{\rm lss} \right|}{G_{\rm N}\rho_{\rm s}} \frac{1}{r_{\rm s}^n} \right)^\frac{1}{1-n},
\end{equation}
Note that this offset diverges for an NFW profile ($n=1$). We also clearly need $n<3$, however this is already required for the halo to have a finite mass within any finite radius. If $n<1$, then the largest offsets will be for the least dense galaxies, whereas the converse is true if $n>1$. The case $n>1$ is interesting in that a larger $\Delta G / G_{\rm N}$ will actually shrink the offset between the BH and galactic centre. However we do not consider this case further since, by considering small perturbations about $r_{\bullet}$, the equilibrium offset is found to only be stable for $n<1$. As mentioned above, $n<1$ is often a better fit for the density profile, so it is not unreasonable to only consider these values.

\begin{figure*}
  \centering
  \includegraphics[width=0.49\textwidth]{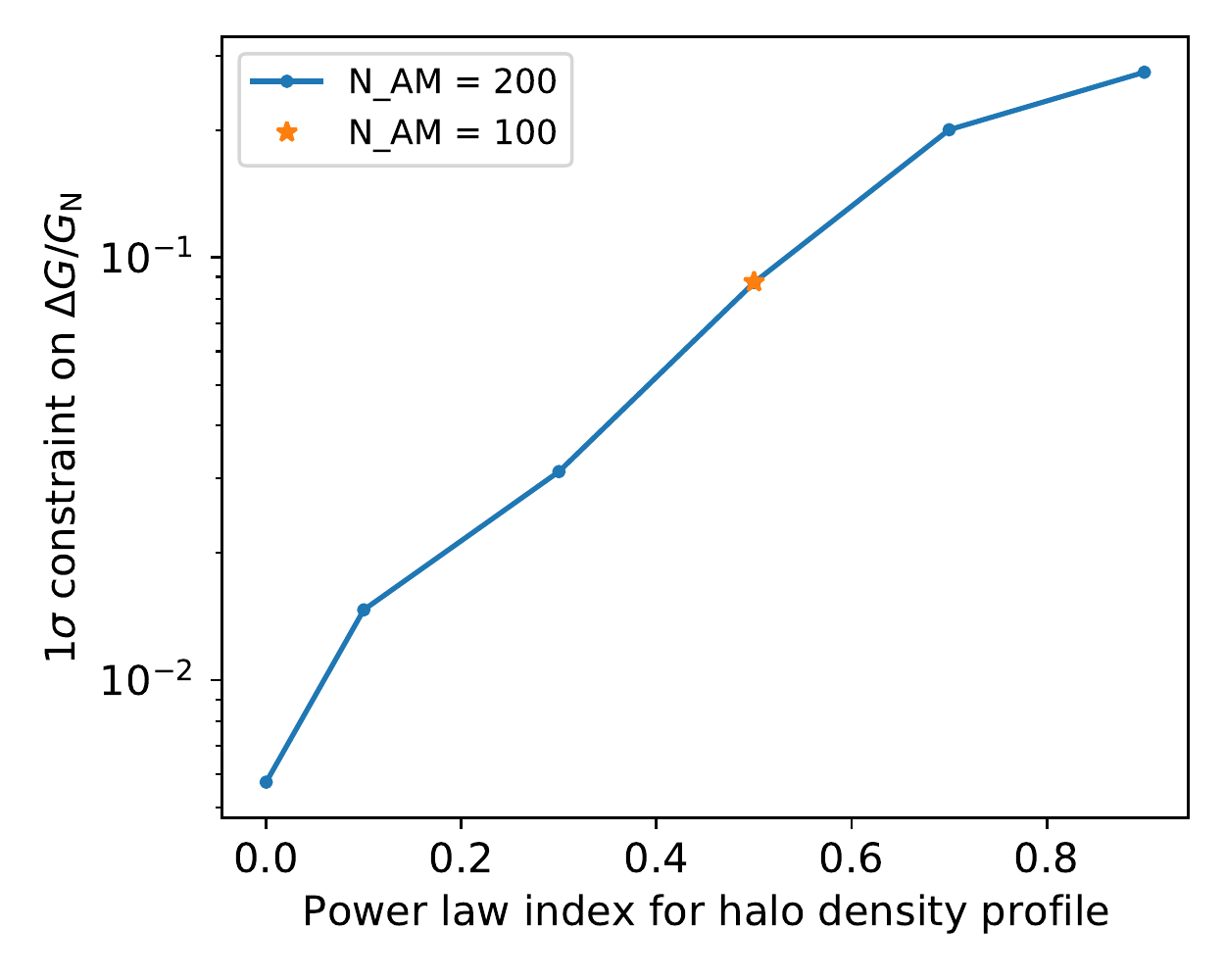}\label{subfig:denspow}
  \includegraphics[width=0.49\textwidth]{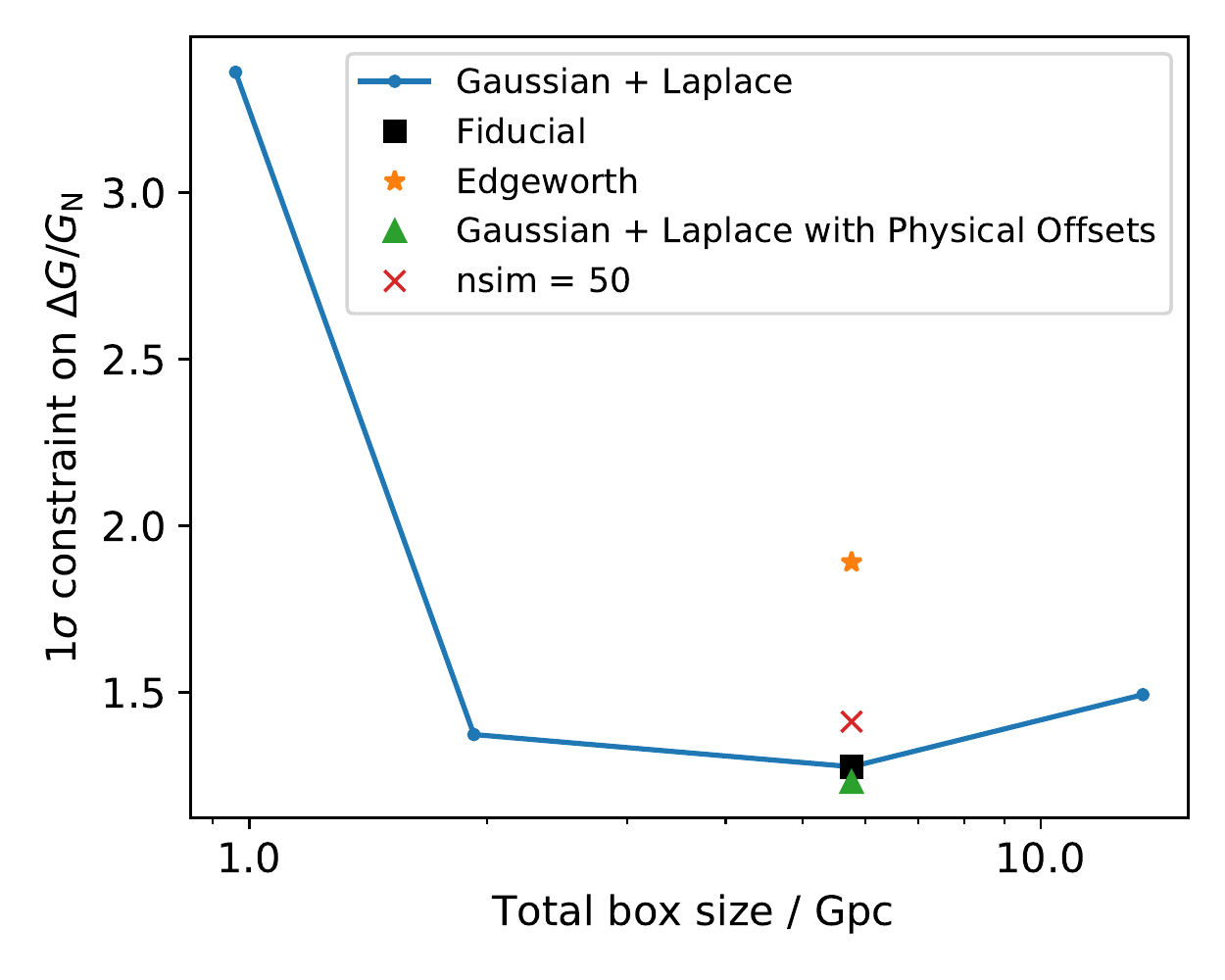}
  \caption{Constraints on $\Delta G / G_{\rm N}$ at $r_{\rm V} = 100 {\rm \, Mpc}$ at $1\sigma$ confidence using the OF13 data.
	 \emph{Left:} A power law halo density profile of index $n$ is assumed.
	 To determine the density profile, we use \texttt{N\_AM}=200 mock catalogues obtained through abundance matching. We re-run the analysis at $n=0.5$ with \texttt{N\_AM}=100 and see that the constraint is unchanged.
	 \emph{Right:} 
	 Using our fiducial density profile, we vary the box size, $L^\prime$, used to add in long wavelength modes when reconstructing the Galileon filed. $L^\prime \sim 1 {\rm \, Gpc}$ corresponds to no extra long wavelength modes. We also show the constraints from different noise models at $L^\prime \sim 6 {\rm \, Gpc}$, and if we reduce the number of constrained simulations to 50.}
	 \label{fig:systematics_plot}
\end{figure*}

The values of $r_{\rm s}$ and $\rho_{\rm s}$ for each test galaxy are estimated using the technique of halo abundance matching (AM) \cite{Conroy, Kravtsov}. The technique assumes a near-monotonic relation between the absolute magnitude of a galaxy and a halo `proxy', typically a combination of virial mass and concentration. We use the best-fit AM model of \cite{Lehmann} applied to the \textsc{rockstar} halo catalogue of the \textsc{DarkSky}-400 N-body simulation \cite{DarkSky} and the S\'{e}rsic $r$-band luminosity function of \cite{Bernardi_SMF}. We generate $\texttt{N\_AM}=200$ mock galaxy catalogues, where each catalogue is a different random realisation of the noise from the intrinsic scatter in the galaxy--halo connection implied by the model. We draw values of $r_{\rm s}$ and $\rho_{\rm s}$ for each galaxy from a randomly chosen catalogue for each Monte Carlo realisation of our model, where we use the halo from that catalogue which is closest in magnitude to that galaxy. By iterating this procedure many times, we marginalise over the stochasticity in the galaxy--halo connection. We re-run the end-to-end inference with $n=0.5$ and only 100 catalogues and find the constraint to be unchanged (\autoref{fig:systematics_plot}), indicating that \texttt{N\_AM} is sufficiently large to sample the distributions of $\rho_{\rm s}$ and $r_{\rm s}$.

As before, we obtain $N_{\rm MC}$ samples from our model of offsets in Galileon gravity and convert this into a GMM for the case $\Delta G / G_{\rm N} = 1$. For a power law profile, the offset is no longer proportional to $\Delta G / G_{\rm N}$, and instead the relation is
\begin{equation} \label{eq:offset_proportionality}
    r_{\bullet} \propto \left( \frac{\Delta G}{G_{\rm N}} \right)^{\frac{1}{1-n}}.
\end{equation}
Thus, to convert the GMM to a different value of $\Delta G / G_{\rm N}$, \autoref{eq:Scaling GMM} is changed to
\begin{equation}
	\tilde{\mu}^{\left( i \right)}_{\rm g, \alpha} = \left( \frac{\Delta G}{G_{\rm N}} \right)^\frac{1}{1-n} \mu^{\left( i \right)}_{\rm g, \alpha}, \quad \tilde{\sigma}^{\left( i \right)}_{\rm g, \alpha} = \left( \frac{\Delta G}{G_{\rm N}} \right)^\frac{1}{1-n} \sigma^{\left( i \right)}_{\rm g, \alpha},
\end{equation}
with the rest of the analysis unchanged from \Cref{sec:Methods}. We again use the Gaussian plus Laplace distribution for our noise model.

In \autoref{fig:systematics_plot} we plot the $1\sigma$ constraints on $\Delta G / G_{\rm N}$ for different power law indices. Fitting the constraint to an exponential, as would be expected from \autoref{eq:power_law_offset}, we find the constraint weakens like $e^{\beta n}$ where $\beta \approx 4.3$. Increasing $n$ increases the density at a given radius which increases the restoring force and thus a larger $\Delta G / G_{\rm N}$ is necessary for a given offset.

Comparing to \autoref{fig:results_corner}, we see that the constraint using a power law profile is an order of magnitude tighter than when we use the scaling relations from \Cref{sec:Calculating the offset} for the smallest values of $n$, and comparable when $n \lesssim 1$. The stronger constraints can be understood in terms of the pivot scale at which the density profile changes from NFW to a power law. In \autoref{eq:power law density profile} we chose $r_{\rm s}$ as the pivot scale however, since we use $n < 1$, if this transition occurred at a smaller radius, the density at a given radius in the power law region would be greater. Changing the pivot scale to $\sim 0.01 r_{\rm s}$ ($\sim 360 {\rm \, pc}$ for a typical galaxy from OF13) would provide constraints similar to our previous prediction. There is no reason \textit{a priori} why we would choose $0.01 r_{\rm s}$ as a pivot scale, but it is reassuring that it is not an unreasonable choice.

Our fiducial case is the halo density profile which gives the most conservative constraint of those considered, hence we report the value derived from scaling relations with a cored profile as our final constraint.

We also assumed that the gas density profile has a S\'{e}rsic index $n_{\rm gas}=1$, i.e. an exponential disk. We repeated the inference with $n_{\rm gas} =0.5$ and $2.0$. Increasing $n_{\rm gas}$ slightly widens the posterior on $\Delta G / G_{\rm N}$. This is expected as increasing $n_{\rm gas}$ increases the central gas density, which decreases the predicted offset for a given $\Delta G / G_{\rm N}$ so slightly larger values of $\Delta G / G_{\rm N}$ are permitted. Given that the gas mass is sub-dominant compared to the other mass components, we would expect the change in the constraint to be small, as indeed it is.

\subsubsection{Noise model}
\label{sec:Systematics - Noise model}

Our previous work on BH offsets \cite{Bartlett_2020} demonstrated that the distribution of offsets is non-Gaussian for these data, so it is unsurprising that the addition of a Laplace distribution is favoured by the BIC. To ensure that our results are insensitive to our choice of non-Gaussianity, we re-run the inference using an Edgeworth expansion for our empirical noise mode, truncating the sum at $F=4$ in \autoref{eq:Edgeworth expansion} as this has the minimum BIC for $3 \leq F \leq 8$. Using the OF13 data, we find that the Edgeworth noise model gives a result which is consistent with zero, with $\Delta G / G_{\rm N} < 1.89$ at $1 \sigma$ confidence, compared to $\Delta G / G_{\rm N} < 1.28$ for the Gaussian plus Laplace distribution with this data.

Although disfavoured by the BIC, we now explore the effect of adding a contribution from physical offsets to the Gaussian plus Laplace distribution noise model (\Cref{sec:Physical Offsets}). We find that our constraint is slightly tightened to $\Delta G / G_{\rm N} < 1.23$ for the OF13 data. As the difference is within the uncertainty on our constraint of 8\% (see \Cref{sec:Results}), we conclude that our constraint is insensitive to this change. We find that $\sigma_{\rm int} < 22 {\rm \, pc}$ at $1 \sigma$ confidence, showing that this contribution is small if it exists at all. We plot the constraints from both of these alternative noise models in \autoref{fig:systematics_plot}.

We conclude that our results are not sensitive to the noise model, provided that we include non-Gaussianity.

\subsubsection{Galileon field}
\label{sec:Systematics: Galileon field}

When reconstructing the Galileon field, one should check that the resolution is sufficiently high such that the maximum $k$ is determined by $r_{\rm V}$ and not the resolution, i.e. that $ L / N < r_{\rm V}$. 

Our minimum $r_{\rm V}$ equals $L/N$ for $N=512$. We have checked that our constraints are unchanged if we use $N=256$ for $r_{\rm V} \geq 10 {\rm \, Mpc}$ (where we require $N > 100$).

Repeating the analysis with 34 constrained simulations did not change the constraint (see \autoref{fig:systematics_plot}), indicating that we have a sufficient number of constrained simulations to sample the distribution of the Galileon field at each galaxy.

In \Cref{sec:Adding larger scale modes} we discussed how we should add Fourier modes to the Galileon field with wavelengths longer than the box length, $L$, of our constrained simulations. This was achieved by creating unconstrained Gaussian random fields in a box of side length $L^\prime = \ell L$. We chose $\ell =6$ (see \Cref{app:Box size for long wavelength modes}), however to check that our constraints are independent of this choice, we re-run the inference with different values of $\ell$ and plot the results in \autoref{fig:systematics_plot}. Comparing $\ell=6$ to $\ell=14$ for the OF13 data, we find the results are consistent, justifying our choice of $\ell=6$. We also find the constraint if we do not include this additional long wavelength information ($\ell=1$) and find that it is weakened from $\Delta G / G_{\rm N} < 1.28$ to $\Delta G / G_{\rm N} < 3.36$ at $1\sigma$ confidence for the OF13 data. This is to be expected as adding in long wavelength modes increases the magnitude of the Galileon field, which tightens the $\Delta G / G_{\rm N}$ constraint.

Removing all of the modes from the constrained simulations is equivalent to setting $r_{\rm V} \sim 1 {\rm \, Gpc}$. We do this in \autoref{fig:DG vs LV} and find little change in our constraints. This is to be expected since the main contributors to the magnitude of the Galileon field are modes with $k \sim 2 \pi / L_{\rm eq}$, so the majority of the modes from the constrained simulation have a negligible impact on this. Even though the constrained simulations turn out to be relatively unimportant in setting our constraints, they would be necessary to obtain a detection because they contain the information on the direction of $\bm{g}$. In the future it will be interesting to repeat the inference using simulations with initial conditions constrained in the much larger SDSS volume \cite{BORG-SDSS}.

\subsubsection{Other potential systematics}

Although we use the BIC to determine how many components to fit in the GMM (\Cref{sec:Gaussian mixture model}), which should penalise components which fit outliers of the distribution, it is important to check that our constraints are not driven by unlikely realisations in our Monte Carlo sampling. Still minimising the BIC, but restricting ourselves to no more than 15 Gaussian components, we find that our constraints are unchanged. We find that all galaxies across all runs require $\sim$10 GMM components even without a maximum number of components, showing that the imposed maximum is not important.

We re-run the end-to-end inference with $\sigma_{\rm D} =1.0, 2.5$ and 5.0${\rm \, dex}$, and find little variation of the constraint with this parameter for the smaller values of $\sigma_{\rm D}$. To understand why increasing the scatter in all of the quantities has little impact on our $\Delta G / G_{\rm N}$ constraint, we fit a single Gaussian instead of a GMM to the distribution of offsets for each galaxy. As expected, the log-normal scatter increases the magnitude of the mean and the width of the Gaussian. Increasing the mean and the covariance have competing effects in $\log\mathcal{L}$ and, until we reach relatively large values of scatter, the two effects cancel and thus the constraint on $\Delta G / G_{\rm N}$ has little dependence on $\sigma_{\rm D}$. Increasing $\sigma_{\rm M}$ to 0.6 or $\sigma_{\rm R}$ to 0.5 also has a negligible impact on our constraint.

The final parameter in the inference is \texttt{ba\_min}, the minimum allowed minor-to-major axis ratio allowed. We set this to 0.15 as this is the lowest axis ratio recorded in the NSA. Changing this to 0.20 was found not to change the constraint.

\subsection{Comparison with the literature}

Previous attempts to constrain Galileons using the polarisation of BH--galaxy offsets relative to the direction of a partially unscreened Galileon field have targeted galaxies in massive galaxy clusters. Since \autoref{eq:cubic Galileon} is a total derivative for a spherically symmetric mass distribution, this equation becomes a modified non-linear Gauss' law; only the mass within some radius sources the field at that point. This results in the Vainshtein mechanism being less efficient inside extended mass distributions, hence why Galileons with sub-{\rm Gpc} values of $r_{\rm C}$ can be constrained in these environments. Constraints $\alpha \lesssim \mathcal{O}(1)$ were obtained \cite{Sakstein_2017,Asvathaman_2017} using the central BH in M87 for Galileons with $r_{\rm C} \lesssim 1 {\rm \, Gpc}$.

By considering galaxies in more rarefied environments, we study the situation where the Galileon field is sourced by large scale structure as opposed to a cluster. This allows us to probe larger values of $r_{\rm C}$, since we no longer require that the mass in the vicinity of the BH sources a partially unscreened Galileon field.

Constraints on Galileons can also be found using the technique of lunar laser ranging \cite{Nordtvedt_1968}, which currently sets the bound \cite{Khoury_2013,Murphy_2012}\footnote{Note that this constraint is sensitive to the rotation vector of the Moon \cite{Hofmann_2018}, which is set to its GR value. This could introduce some model dependence in the result.}
\begin{equation}
    r_{\rm C} \alpha^{-\frac{3}{2}} \gtrsim 150 {\rm \, Mpc}.
\end{equation}
Slightly weaker constraints are obtained by studying the precession of planetary orbits in the Solar System \cite{Battat_2008}.

\begin{figure}
	\centering
	\includegraphics[width=\columnwidth]{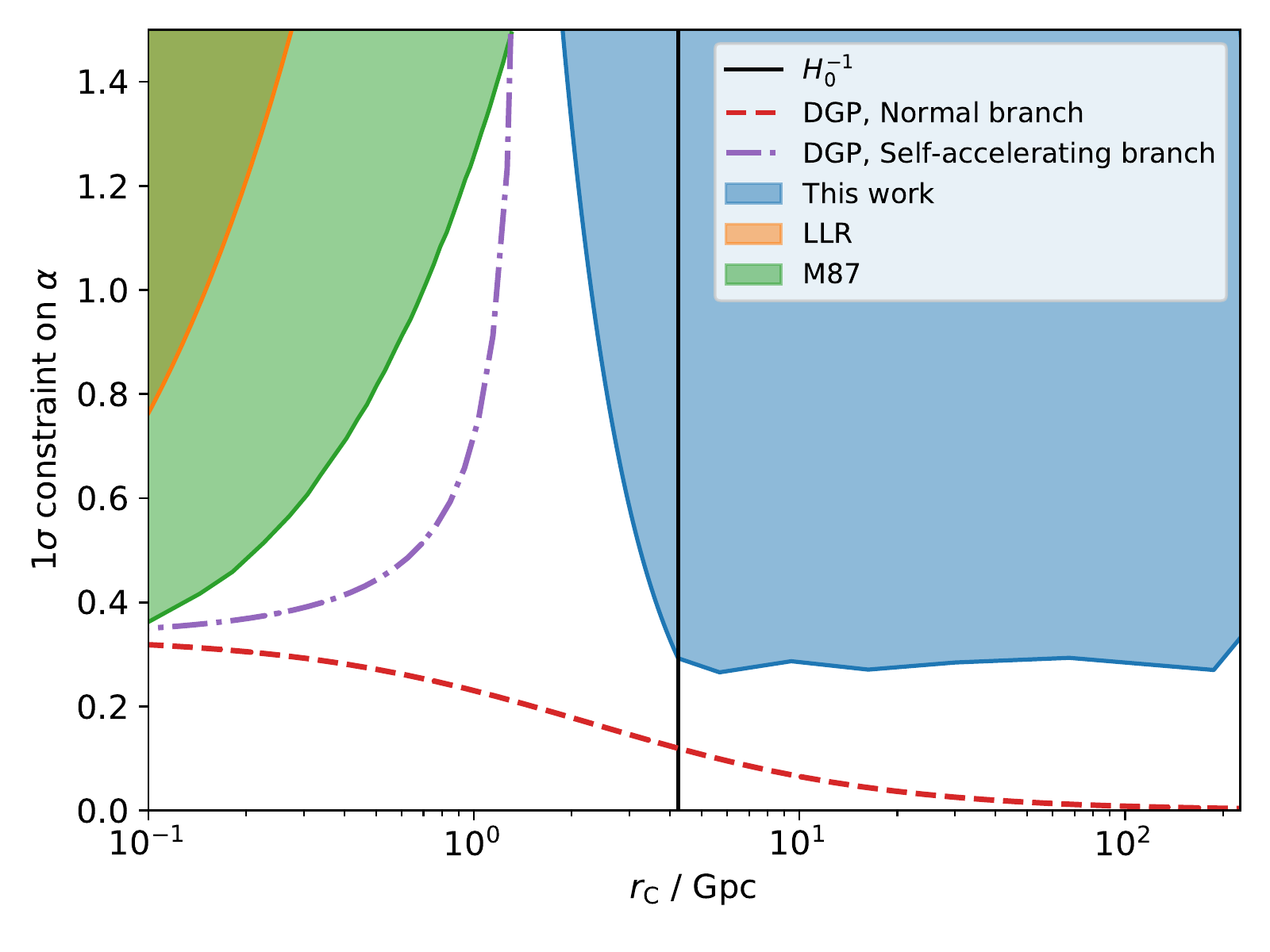}
	\caption{\label{fig:alpha-rc constraint} Constraints on the coupling of a cubic Galileon to matter, $\alpha$, as a function of the crossover scale, $r_{\rm C}$, from lunar laser ranging (LLR) \cite{Khoury_2013}, the BH at the centre of M87 \cite{Sakstein_2017,Asvathaman_2017} and this work. The shaded regions are excluded at $1\sigma$ confidence. We also plot the $\alpha-r_{\rm C}$ curves for the normal and self-accelerating branches of the DGP model (\autoref{eq:alpha DGP}).}
\end{figure}

In \autoref{fig:alpha-rc constraint} we plot the constraints on $\alpha$ as function of $r_{\rm C}$ for the cubic Galileon from the literature and compare to the constraints obtained in this work. As discussed in \Cref{sec:Modelling the Galileon field}, we convert $r_{\rm V}$ to $r_{\rm C}$ by using \Cref{eq:Cosmological mass enclosed,eq:rv to rc conversion}. From this we see that our constraints are applicable to the region $r_{\rm C}\gtrsim H_0^{-1}$. As anticipated, this is complementary to previous work, which constrains $r_{\rm C} \sim 0.01-1 {\rm \, Gpc}$, and is comparable in strength.

Our conversion from $r_{\rm V}$ to $r_{\rm C}$ in a cosmological context is based on the non-linear matter power spectrum from \texttt{CLASS}; for smaller values of $r_{\rm V}$ within the 1-halo term this approximation breaks down and an alternative conversion would be required. We anticipate that if we were to improve the modelling of the Galileon field to incorporate the non-linear regime, our constraint on $\alpha$ would remain relatively unchanged as we moved to smaller $r_{\rm C}$. Our test will then become competitive with Lunar Laser Ranging and constrain self-accelerating Galileons, providing an alternative to the Integrated Sachs-Wolfe probe \citep{Renk_2017}.

We note that the strength of our constraints are similar to those from M87 \cite{Sakstein_2017}, despite our sample containing 1916 galaxies and their one. This is due to the interplay of three effects. First, the observations used in this work have lower resolution, with $\sigma \sim 50 {\rm \, mas}$ for the radio data, whereas the galaxy--BH offset for M87 is measured to be $< 30 {\rm \, mas}$. Second, we marginalise over an empirical noise model, whereas \cite{Sakstein_2017} assume that the entire offset is due to a fifth force, which would make their constraint tighter but also more prone to systematics to do with astrophysical contributions to the offset. Finally, we only consider the Galileon field sourced by large scale structure, which is smaller than that near a massive cluster and hence allows larger $\Delta G / G_{\rm N}$ values for a given offset. Combining cosmic and cluster fields in future work will therefore afford much tighter constraints.

\begin{figure}
	\centering
	\includegraphics[width=\columnwidth]{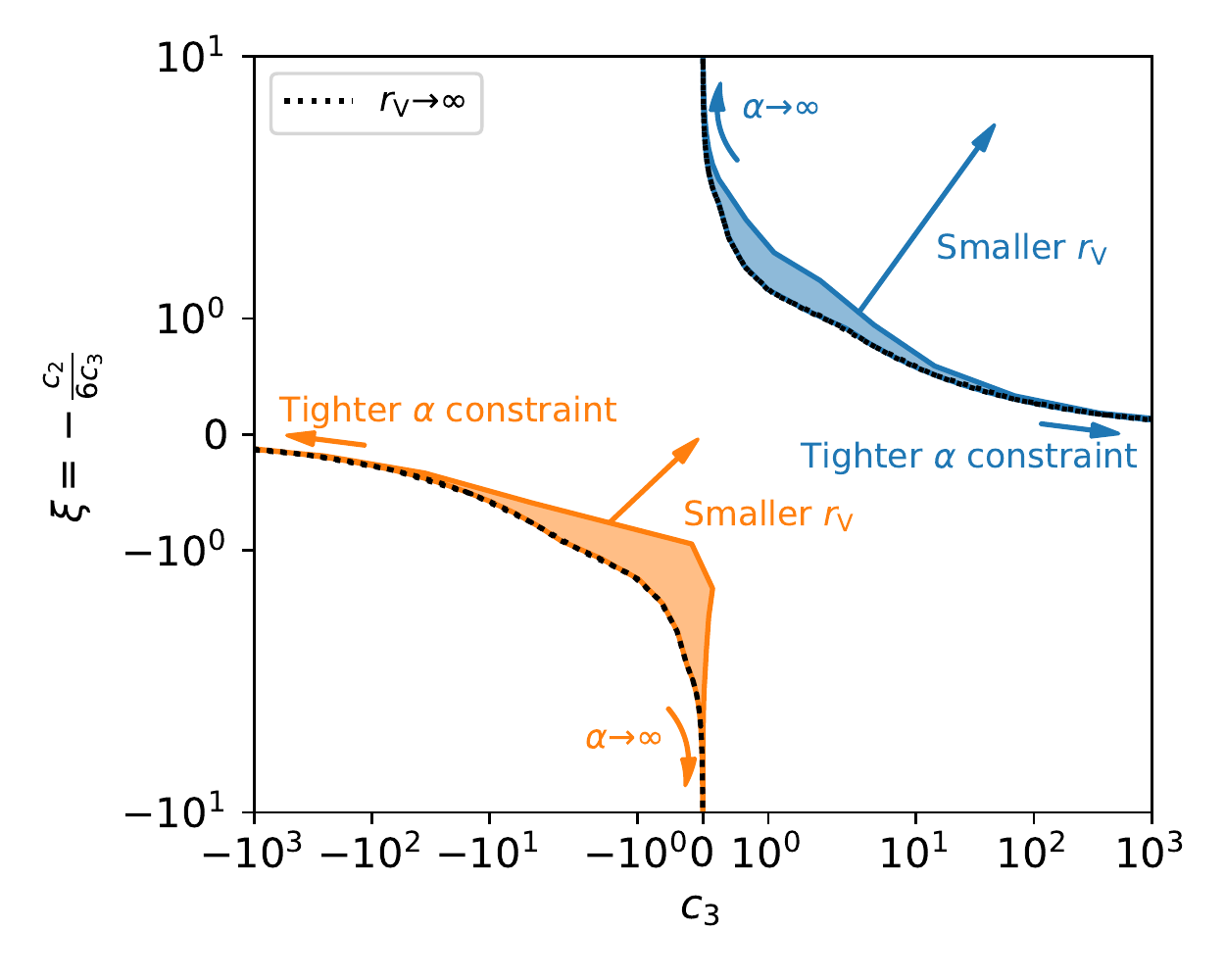}
	\caption{\label{fig:c3-xi constraint}Constraints on the Horndeski parameters $c_2$ and $c_3$ for the cubic Galileon tracker solution $\dot{\bar{\varphi}}H = \xi H_0^2$ at $1 \sigma$ confidence. The blue region corresponds to $\xi >0$ and the orange region is for $\xi < 0$. Lines of constant $r_{\rm V}$ are given by the curves $\xi^3 c_3 = {\rm constant}$, where $\xi$ is proportional to $\alpha$ along these curves. Probing smaller $r_{\rm V}$ in future work will push the constraint to the upper right, as shown by the arrows.}
\end{figure}

To enable easy comparison to the literature, we also convert our constraint on $r_{\rm V}$ and $\Delta G / G_{\rm N}$ to one on the Horndeski parameters $c_2$ and $c_3$. To do this, we consider the tracker solution \cite{deFelice_2010}
\begin{equation}
    \dot{\bar{\varphi}} H = \xi H_0^2 = {\rm constant},
\end{equation}
where $\xi$ is related to the dimensionless Horndeski parameters for cubic Galileons as
\begin{equation}
    \xi = - \frac{c_2}{6 c_3}.
\end{equation}
From \Cref{eq:cubic Galileon Barreira,eq:beta1 definition,eq:beta2 definition,eq:cubic Galileon} we see that today
\begin{equation}
   \beta_1 = \frac{\xi}{3} \left[ c_3 \xi^3 - 1 + 2 \frac{\dot{H}}{H_0^2} \right], \quad
    \beta_2 = \frac{2}{\xi^2} \beta_1,
\end{equation}
where
\begin{equation}
    \beta_1 = \left( H_0 r_{\rm C} \right)^{-2}, \quad \beta_2 = \frac{1}{3\alpha}.
\end{equation}
From this conversion and \autoref{eq:rv to rc conversion}, we see that lines of constant $r_{\rm V}$ are transformed to lines of constant $c_3 \xi^3$, where $\xi$ is proportional to $\alpha$ on these curves. In \autoref{fig:c3-xi constraint} we plot our constraints in the $c_3-\xi$ plane. In this plot we demonstrate the regions of parameter space that can be probed by our test, and ways of further constraining this region. It is clear from the plot that further work should target smaller $r_{\rm V}$, as our constraint already lies close to the line corresponding to $r_{\rm V} \to \infty$.

We also plot the $\alpha-r_{\rm V}$ curves in \autoref{fig:alpha-rc constraint} for the normal and self-accelerating branches of the DGP model, evaluated using \autoref{eq:alpha DGP} and assuming the matter density and $H_0$ is the same as in the constrained simulations. We see that the normal branch is not yet constrained by our test. For our smallest value of $r_{\rm V}$, we would require a constraint $\alpha < 0.04$ to do this, corresponding to an improvement of a factor of $\sim 7$. 

Due to the assumption of linearity, our constraints are insensitive to the specific Galileon model; using the Vainshtein radius for a quartic Galileon in \autoref{fig:alpha-rc constraint} changes the conversion to $r_{\rm C}$ by a numerical factor of $\mathcal{O}(1)$ \cite{Novel_Probes_2019}. Our bounds are therefore equally applicable to quartic and quintic Galileons. The results of GW170817 \cite{GW170817} already severely constrain the self-accelerating branches of these models, with constraints on the Horndeski parameters \cite{Ezquiaga_2017}
\begin{equation}
    |c_4| \lesssim 2.8 \times 10^{-17} \left( \frac{2}{\xi} \right)^4, \quad |c_5| \lesssim 3.8 \times 10^{-17} \left( \frac{2}{\xi} \right)^5,
\end{equation}
where $\xi = H(t)\dot{\varphi}/H_0^2$. To convert these to bounds on $\alpha$ and $r_{\rm C}$ using variants of \Cref{eq:beta1 definition,eq:beta2 definition} relevant to these models, one would also need to know $c_3$. Since this is not constrained by GW170817, we do not perform this comparison explicitly, but note that our results provide independent stringent constraints on the quartic and quintic models.

\vfill\eject

\section{Conclusions}
\label{sec:Conclusions}

Galileons are scalar field theories which obey the Galileon symmetry $\varphi \to \varphi + b + c_\mu x^\mu$ \cite{Nicolis_2009} and are parametrised by their crossover scale, $r_{\rm C}$, and coupling to matter, $\alpha$. If $r_{\rm C} \sim H_0^{-1}$ then the Galileon is said to be cosmologically relevant and could explain the late-time accelerated expansion of the Universe. Here we present a test of Galileon gravity by comparing the predicted equilibrium offsets between the centre of a galaxy and its central supermassive BH to observational data.

In a similar manner to recent work which ruled out astrophysically relevant $f(R)$ theories \cite{f(R)_ruled_out}, we construct a galaxy-by-galaxy Bayesian forward model of these offsets based on dynamical information about these galaxies, a reconstruction of the local gravitational field and a non-Gaussian empirical noise model. In doing so we introduced \texttt{CSiBORG}, a suite of constrained N-body simulations of the local Universe using initial conditions from the \texttt{BORG} algorithm. Marginalising over noise parameters and propagating uncertainties on input quantities via Monte Carlo sampling, we derive constraints on the magnitude of the fifth force arising from cosmologically relevant Galileons, $\Delta G / G_{\rm N} \equiv 2 \alpha^2$.

We rule out $\Delta G / G_{\rm N} > 0.16$ at $1 \sigma$ confidence for $r_{\rm C} \gtrsim H_0^{-1}$. We find our constraints to be robust to the assumed halo density profile, the choice of empirical noise model and parameters used to infer physical properties from the dynamical information available on these galaxies.  These constraints are complementary to previous constraints from galaxy--BH offsets \cite{Asvathaman_2017,Sakstein_2017}: we probe larger values of $r_{\rm C}$ because we consider the Galileon field sourced by large scale structure as opposed to a massive galaxy cluster.

Improved modelling of the Galileon field to enter the non-linear regime should make our constraints competitive with lunar laser ranging (LLR) \cite{Khoury_2013} at smaller values of $r_{\rm C}$. Our approach necessarily requires a large number of galaxies to fit the parameters of the empirical noise model accurately, so targeted observations at a small number of galaxies would not be particularly useful in tightening our constraints. Improved modelling of dynamical friction in future cosmological hydrodynamical simulations \cite{Bartlett_2020} could remove our reliance on an empirical noise model, making such simulations useful for future constraints on Galileons and similar theories.

\acknowledgements
{
We thank Alex Barreira, Emilio Bellini, Lam Hui, Kazuya Koyama, Jeremy Sakstein, Hans Winther, and Miguel Zumalacarregui for useful discussions, and Julien Devriendt, Mladen Ivkovic, Jens Jasche, Guilhem Lavaux and Adrianne Slyz for input on the constrained simulations. We are grateful to Scott Barrows, Sandor Frey and Gabor Orosz for providing data necessary for this work.

DJB is supported by STFC and Oriel College, Oxford. HD is supported by St John's College, Oxford, and acknowledges financial support from ERC Grant No 693024 and the Beecroft Trust. PGF is supported by the ERC, STFC, and the Beecroft Trust.

This work used the DiRAC Complexity system, operated by the University of Leicester IT Services, which forms part of the STFC DiRAC HPC Facility (www.dirac.ac.uk). This equipment is funded by BIS National E-Infrastructure capital grant ST/K000373/1 and STFC DiRAC Operations grant ST/K0003259/1. DiRAC is part of the National E-Infrastructure.
}

\bibliographystyle{apsrev4-1}
\bibliography{references}

\appendix

\section{Box size for long wavelength modes}
\label{app:Box size for long wavelength modes}

When adding in long wavelength modes to account for scales not captured by the constrained simulations, we should ensure that the size of our larger box, $L^\prime = \ell L$, is sufficiently large to accurately capture all small-$k$ modes. To find a suitable value for this, for simplicity we assume that the modes are continuous (so we are integrating rather than summing on a grid) and denote these by the subscript `$\rm cts$'. In this case, the expectation value of the square of the gravitational field at the origin is
\begin{equation} \label{eq:Mean g^2}
    \begin{split}
        \left< \left| \bm{g}_{\rm cts} \right|^2 \right> &=  \left( 4 \pi G \bar{\rho} \right)^2 \lim_{r \to 0} \int \frac{{\rm d}^3 \bm{k}}{\left( 2 \pi \right)^3} \frac{P \left( k \right)}{k^4} e^{2 i \bm{k} \cdot \bm{r}} \\
        & = \left( 2 G \bar{\rho} \right)^2 \lim_{r \to 0} \int_0^\infty P \left( k \right) \frac{\sin \left( 2 k r \right)}{kr} {\rm d}k,
    \end{split}
\end{equation}
since the power spectrum is defined to be
\begin{equation}
    \left< \Delta_{\rm cts} \left( \bm{k} \right) \Delta_{\rm cts}^* \left( \bm{k}^\prime \right) \right> \equiv \left( 2\pi \right)^3 P \left( k \right) \delta^{\rm D} \left( \bm{k} + \bm{k}^\prime \right),
\end{equation}
where $\delta^{\rm D}$ is the Dirac delta function. For simplicity, we assume the matter power spectrum can be described by a broken power law
\begin{equation}
	P \left( k \right) \approx 
  \begin{cases}
    P_{\rm eq} \left( \frac{k}{k_{\rm eq}} \right), & k < k_{\rm eq} \\
    P_{\rm eq} \left( \frac{k}{k_{\rm eq}} \right)^{-3}, & k > k_{\rm eq}.
  \end{cases}
\end{equation}

For a finite box size, we cannot calculate the integral in \autoref{eq:Mean g^2}, since we can only integrate from some finite $k_0$ to obtain $\langle|\bm{g}_{\rm cts}|^2 \rangle_0$, such that
\begin{equation}
    \left< \left| \bm{g}_{\rm cts} \right|^2 \right> = \left< \left| \bm{g}_{\rm cts} \right|^2 \right>_0 + \delta g_{\rm cts}^2 \left( k_0 \right),
\end{equation}
where our correction from long wavelength modes is
\begin{equation} \label{eq:g_cts correction definition}
	\delta g_{\rm cts}^2  \left( k_0 \right) \equiv \left( 2 G \bar{\rho} \right)^2  \lim_{r \to 0} \int_0^{k_0} P \left( k \right) \frac{\sin \left( 2 k r \right)}{kr} {\rm d}k.
\end{equation}
In the case where $k_0 < k_{\rm eq}$ we have 
\begin{equation}
	\delta g_{\rm cts}^2 =  \left( 2 G \bar{\rho} k_0 \right)^2 \frac{ P_{\rm eq}}{k_{\rm eq}},
\end{equation}
whereas for $k_0 > k_{\rm eq}$ we have 
\begin{equation}
    \begin{split}
        \delta g_{\rm cts}^2 &= \delta g_{\rm cts}^2 \left( k_{\rm eq} \right) +  \left( 2 G \bar{\rho} \right)^2  \lim_{r \to 0} \int_{k_{\rm eq}}^{k_0} P \left( k \right) \frac{\sin \left( 2 k r \right)}{kr} {\rm d}k \\
        &=  \left( 2 G \bar{\rho} \right)^2 P_{\rm eq} k_{\rm eq} \left( 2 - \left( \frac{k_{\rm eq}}{k_0} \right)^2 \right).
    \end{split}
\end{equation}
By sending $k_0 \to \infty$, we see that
\begin{equation}
	\left< \left| \bm{g}_{\rm cts} \right|^2 \right> =  8 \left( G \bar{\rho} \right)^2 P_{\rm eq} k_{\rm eq},
\end{equation}
and so we arrive at, defining $L_0 \equiv 2 \pi / k_0$ and $L_{\rm eq} \equiv 2 \pi / k_{\rm eq}$,
\begin{equation} \label{eq:g_cts correction result}
	\frac{\delta g_{\rm cts}^2 \left( L_0 \right)}{\left< \left| \bm{g}_{\rm cts} \right|^2 \right>} =
	\begin{cases}
    \frac{1}{2} \left( \frac{L_{\rm eq}}{L_{0}} \right)^{2}, & L_0 > L_{\rm eq} \\
    1 - \frac{1}{2} \left( \frac{L_{0}}{L_{\rm eq}} \right)^{2}, & L_0 < L_{\rm eq}.
  \end{cases}
\end{equation}
Using $L_{\rm eq} \approx 450 {\rm \, Mpc}$, we find that for $L^\prime = 6 {\rm \, Gpc}$ ($\ell \approx 6$), the correction is $\sqrt{\delta g_{\rm cts}^2 / <|\bm{g}_{\rm cts}|^2>} \approx 0.05$, or for $L^\prime = 14 {\rm \, Gpc}$ ($\ell \approx 14$), this is $\approx 0.02$. This is small compared to the $\sim30$ per cent variation in $|\bm{g}|$ across the constrained simulations, so using either of these values is acceptable.

\section{Coefficients in the Edgeworth expansion}
\label{app:Coefficients in the Edgeworth expansion}

The non-zero coefficients $\alpha_n$ in the Edgeworth expansion (\autoref{eq:Edgeworth expansion}) are (up to $F=8$)
\begin{equation}
	\begin{split}
	& \alpha_0 = 1, \quad \alpha_3 = \frac{\kappa_3}{2^{3/2} \times 3!},\quad \alpha_4 = \frac{\kappa_4}{2^2\times 4!}, \\
	&\alpha_5 = \frac{\kappa_5}{2^{5/2} \times 5!}, \quad \alpha_6 = \frac{10 \kappa_3^2 + \kappa_6}{2^3 \times 6!} \\
	& \alpha_7 = \frac{35 \kappa_3  \kappa_4 + \kappa_7}{2^{7/2} \times 7!}, \quad \alpha_8 = \frac{35 \kappa_4^2 + 56 \kappa_3 \kappa_5 + \kappa_8}{2^4 \times 8!} ,
	\end{split}
\end{equation}
where $\kappa_n$ is the $n^{\rm th}$ cumulant. The powers of $2^{n/2}$ are due to us using the Physicist's and not the Statistician's Hermite Polynomials. For notational convenience, we define the following parameters
\begin{equation}
    \label{eq:Define Edgeworth parameters}
    \begin{split}
	&\gamma \equiv \frac{\kappa_3}{2^{3/2}}, \quad \tau  \equiv \frac{\kappa_4}{2^{2}}, \quad \eta \equiv \frac{\kappa_5}{2^{5/2}} \\
	&\zeta \equiv \frac{\kappa_6}{2^{3}}, \quad \xi \equiv \frac{\kappa_7}{2^{7/2}}, \quad \iota \equiv \frac{\kappa_8}{2^{4}},
	\end{split}
\end{equation}
so the coefficients are
\begin{equation}
	\begin{split}
	& \alpha_0 = 1, \quad \alpha_3 = \frac{\gamma}{3!}, \quad \alpha_4 = \frac{\tau}{4!}, \quad \alpha_5 = \frac{\eta}{5!}, \\
	& \alpha_6 = \frac{1}{6!} \left( \zeta + 10 \gamma^2 \right), \quad \alpha_7 = \frac{1}{7!} \left( \xi +35 \gamma \tau\right)\\
	& \alpha_8 = \frac{1}{8!} \left( \iota + 56 \gamma \eta + 35 \tau^2 \right).
	\end{split}
\end{equation}
We need to impose restrictions on the parameters since they must describe a probability distribution. Following \cite{Dubkov_1976}, we define
\begin{equation}
	\left< a, b \right> \equiv \left< a b \right> - \left< a \right> \left< b \right>,
\end{equation}
and, since the probability density is non-negative, we use the Cauchy-Bunyakovskii Inequality
\begin{equation}
	\left< a, b \right>^2 \leq \left< a^2 \right> \left< b^2 \right>,
\end{equation}
which for some random variable $x$ gives
\begin{equation}
\label{eq:Cauchy-Bunyakovskii}
	\left< x^p, x^q \right>^2 \leq \left< x^p, x^p \right> \left< x^q, x^q \right>.
\end{equation}
These can be trivially expressed in terms of cumulants, and thus we have necessary conditions for the parameters $\bm{\Omega}=\{\sigma_{\rm obs},\gamma,\tau,\eta,\zeta,\xi,\iota\}$ to describe a probability distribution. We impose this constraint for all $p < q \leq 4$ and consider the cases $3 \leq F \leq 8$.

\end{document}